\documentclass[twocolumn,traditabstract]{aa}

\usepackage{natbib} \usepackage{subcaption}
\usepackage{amssymb,amsfonts} \usepackage{amsmath,mathtools}
\usepackage{url} \usepackage{hyperref} \usepackage{xcolor}
\usepackage{graphics}
\usepackage[flushleft]{threeparttable}

\newcommand{\ha}{$\mathrm{H\alpha}~$}
\newcommand{\fig}{Fig. }
\newcommand{\figs}{Figs. }
\newcommand{\sect}{Sect. }

\begin{document}

\title{Spectropolarimetric observations of the solar atmosphere \\in
  the \ha 6563\,\r{A} line}

\author{J. Jaume Bestard\inst{1,2} \and
J. Trujillo Bueno\inst{1,2,3} \and
M. Bianda\inst{4} \and
J. \v{S}t\v{e}p\'an\inst{5} \and
R. Ramelli\inst{4}}

\institute{ Instituto de Astrof\'{\i}sica de Canarias, V\'{\i}a
  L\'actea s/n, E-38205 La Laguna, Tenerife, Spain.
  \and Departamento de Astrof\'{\i}sica,
  Universidad de La Laguna (ULL), E-38206 La Laguna, Tenerife, Spain
  \and Consejo Superior de Investigaciones Cient\'{\i}ficas,
  Spain. \email{jtb@iac.es} \and Istituto Ricerche Solari (IRSOL),
  Università della Svizzera italiana (USI), CH-6605 Locarno-Monti,
  Switzerland. \and Astronomical
  Institute ASCR, v.v.i., Ond\v{r}ejov, Czech
  Republic.}

\date{Accepted 16/12/2021}

\abstract{We present novel spectropolarimetric observations of the
hydrogen \ha line taken with the Z\"urich Imaging Polarimeter (ZIMPOL)
at the Gregory Coud\'e Telescope of the Istituto Ricerche Solari
Locarno (IRSOL). The linear polarization is clearly dominated by the
scattering of anisotropic radiation and the Hanle effect, while the
circular polarization by the Zeeman effect. The observed linear
polarization signals show a rich spatial variability, the
interpretation of which would open a new window for probing the solar
chromosphere.  We study their spatial variation within coronal holes,
finding a different behaviour for the $U/I$ signals near the North and
South solar poles. We identify some spatial patterns, which may
facilitate the interpretation of the observations.  In
close-to-the-limb regions with sizable circular polarization signals
we find similar asymmetric $Q/I$ profiles. We also show examples of
net circular polarization profiles (NCP), along with the corresponding
linear polarization signals. The application of the weak field
approximation to the observed circular polarization signals gives
$10\,$G ($40-60\,$G) in close to the limb quiet (plage) regions for
the average longitudinal field strength over the spatio-temporal
resolution element.}


\titlerunning{Spectropolarimetric observations in the \ha line}
\authorrunning{J. Jaume Bestard et al.}

\maketitle


\section{Introduction\label{sec:intro}}

The \ha line is often used to investigate the solar chromosphere and
its energetic events, like filaments, Ellerman bombs, surges and
flares \cite[][]{doi:10.1146/annurev-astro-081817-052044}. These
events leave some spectral signatures in the \ha intensity profiles,
which can be used for the target classification and for studying the
fine-scale structure and temporal evolution of the chromosphere.  In
the last decades, thanks to the increasing spatial, spectral, and
temporal resolution of the new instruments, it has been possible to
investigate many of such events through their \ha intensity profiles
\cite[e.g.][]{2019ApJ...875L..18S, 2019A&A...626A...4V,
2021ApJ...907...54V}.

However, little has been done concerning the polarization of the \ha
line. This is due to the fact that its polarization signals are hard
to measure and difficult to interpret, especially the linear
polarization caused by scattering processes
\citep[see][]{2010ApJ...711L.133S,2011ApJ...732...80S}.  The \ha line
is composed of 7 overlapped radiative transitions among its 11
fine-structure (FS) levels, and it is sensitive to depolarizing
collisions with protons and electrons at the chromospheric densities
and temperatures \cite[][]{1996A&A...309..317S}. Some investigations
have reported spectropolarimetric observations of \ha in flares
\cite[e.g. ][]{1999A&A...349..283V, 2005A&A...434.1183B,
2007A&A...465..621S, 2013A&A...556A..95H}, Ellerman Bombs
\cite[e.g][]{2002ESASP.506..661K, 2006SoPh..235...75S}, prominences
\cite[e.g.][]{2003AN....324..324K, 2005ApJ...621L.145L} and other
events in the solar atmosphere. However, observations aimed at
investigating the \ha polarization in quiet regions are very scarce.
\cite{1997A&A...322..985S} and \cite{2000A&A...355.1138C} investigated
the center-to-limb variation (CLV) of the $Q/I$ scattering
polarization signal. In addition, the atlas of the ``second solar
spectrum'' \cite[][]{2000sss..book.....G} shows an interesting
asymmetric $Q/I$ profile with an amplitude of $0.12\,\%$. Some
theoretical investigations have improved our understanding about this
spectral line in quiet solar regions, both concerning the line's
intensity and the polarization.  \cite{2004ApJ...603L.129S} studied
the response function of the intensity and circular polarization in
one-dimensional (1D) models, concluding that only the line-core is
sensitive to perturbations in the chromosphere. However, through
radiative transfer simulations of its Stokes $I$ profile,
\cite{2012ApJ...749..136L} investigated the formation of the \ha line
in a three-dimensional (3D) magnetohydrodynamical (MHD) model,
concluding that the line-core forms almost always in the low-beta
plasma regime, thus supporting the idea that the \ha line is a good
tracer of the chromospheric plasma. Moreover, they compared
calculations assuming 1D and 3D (see their \fig 7),
revealing very different images at the center of the \ha
line. \cite{2010ApJ...711L.133S,2011ApJ...732...80S} studied in detail
the scattering polarization and Hanle effect of the \ha line in 1D
models, considering the role of collisions with protons and
electrons. By means of 1D radiative transfer calculations, these
authors concluded that via the Hanle effect the line's scattering
polarization is sensitive to the presence of magnetic field gradients
in the upper chromosphere.

The present paper aims at improving our observational knowledge of the
\ha polarization signals that scattering processes and the Hanle and
Zeeman effects produce in quiet and plage regions of the solar
disk. We present novel spectropolarimetric observations, in which the
spatial and polarimetric resolutions are sufficient to detect a high
spatial variability in the emergent polarization profiles. Different
solar regions near the limb are analyzed, namely coronal holes, quiet,
and plage regions. \sect \ref{sec:observations} describes the used
instrumentation, the observations and the reduction process, while in
\sect \ref{sec:results} we analyze the reduced data. \sect
\ref{subsec:averaged} focuses on the CLV of the polarization signals,
both the line-center amplitudes and the shape of the profiles. The
spatial variations of the polarization signals along the slit are
shown in \sect \ref{subsec:variation}. \sect \ref{subsec:net} shows
examples of net circular polarization (NCP) signals, while in \sect
\ref{subsec:wfa} we apply the weak field approximation (WFA). Finally,
in \sect \ref{sec:conclusions} we summarize our main conclusions and
discuss avenues for future research.


\section{Observations and data reduction\label{sec:observations}}


\subsection{Instrumentation \label{ssec:instr}}

The Gregory-Coud\'e telescope at IRSOL has a 45 cm aperture and 25
meters of effective focal length. The two off-axis plane mirrors
placed after the secondary mirror deflect the beam into the
declination and hour axes, respectively. The relative orientation of
these two mirrors depends on the declination and, consequently, a
practically constant instrumental polarization due to oblique
reflections is obtained during the day \citep{1991SoPh..134....1A}.

The Czerny-Turner echelle spectrograph has $10\,$m of focal length and
uses a $180\times 360$ mm grating with 316 lines per mm and $63^{\circ}$
blaze angle. A set of high transmission pre-filters installed on a
filter wheel allows to choose the spectral range of the light entering the
spectrograph, avoiding the overlap of different grating orders.

The main advantage of the Z\"urich IMaging POLarimeter (ZIMPOL) is
that it is able to operate at a high modulation frequency, in our case
42 kHz that corresponds to the eigenfrequency of the piezo-elastic
modulator (PEM) used. The ZIMPOL camera is equipped with an advanced
CCD sensor on which three out of four pixel rows are masked, while
cylindrical micro-lenses focus the light on the free pixel rows. By
shifting the photo-charges synchronously with the modulator and using
the masked pixel rows as buffer, it is possible to get four intensity
images acquired during four different phase intervals of the
modulation. This allows to demodulate the light signal and to retrieve
the Stokes images. The fast modulation frequency implies that
seeing-induced cross-talks are suppressed and then it allows us to
achieve an unprecedented polarimetric sensitivity of about $10^{-5}$
with long exposure times \cite[][]{10.1117/12.857120}. In order to
measure the full Stokes vector with the PEM modulator, we carried out
two independent measurements: first $I$, $Q/I$, and $V/I$ and then,
after rotating the analyzer by $45^{\circ}$, $I$, $U/I$, and $V/I$.

The spectral and spatial dimensions are covered by 1240$\times$140
pixels. The width and length of the slit subtend $0.5\arcsec$ and
$200\arcsec$ on the solar disk, respectively.  This gives us a spatial
sampling of $1.4\arcsec$/pixel in the spatial direction. The acquired
spectral images cover a range of $10.8\,$\r{A} with a spectral
sampling of $9.7\,$m\r{A}/pixel.


\subsection{The spectropolarimetric observations\label{ssec:reduction}}

The observational campaign of 5 days took place from May 29 to June 2
of 2019. The main goal was to measure the linear and circular
polarization signals of \ha in quiet and active regions. We took some
measurements at different limb distances in order to determine the
center-to-limb variation (CLV). The slit was always placed parallel to
the nearest solar limb. Table \ref{tab:observations} collects all the
observations of scientific interest of the campaign. The observations
at the West limb were taken in a quiet region while the observations
at the North and South limbs in coronal holes. The observations at the
East limb were taken in a plage region. The table indicates that the
typical noise per pixel in the polarization signals is slightly above
$10^{-4}$ for 9 minutes of exposure time. The slit position is
controlled by the Primary Guiding System (PIG)
\citep{1998SoPh..182..247K,2011AN....332..502K}, with a precision of
about $2 \arcsec$. Thanks to a limb tracking system based on a glass
tilt plate, the slit position was kept stable with respect the solar
limb allowing a precision of about $0.5\arcsec$.

The reduction process of the data had the following steps: 1)
demodulation of the raw measurements in order to recover the Stokes
images, 2) correction of the flat-field and data measurements from the
dark current, 3) application of the polarimetric calibration to the
data, 4) correction of the intensity image from the flat-field, 5)
removal of fringes through a FFT filter to the polarization images, 6)
correction of cross-talks $V\rightarrow Q,U$ using empirical
measurements, as reported by \cite{2005ESASP.596E..82R}, and 7) we
substract the continuum polarization because we are only interested in
the variation with wavelength of the spectral line polarization.


\section{Results\label{sec:results}}

\subsection{Spatially averaged profiles\label{subsec:averaged}}

\begin{figure*}[h!]\centering
  \begin{subfigure}[b]{0.82\textwidth}
    \includegraphics[width=\columnwidth]{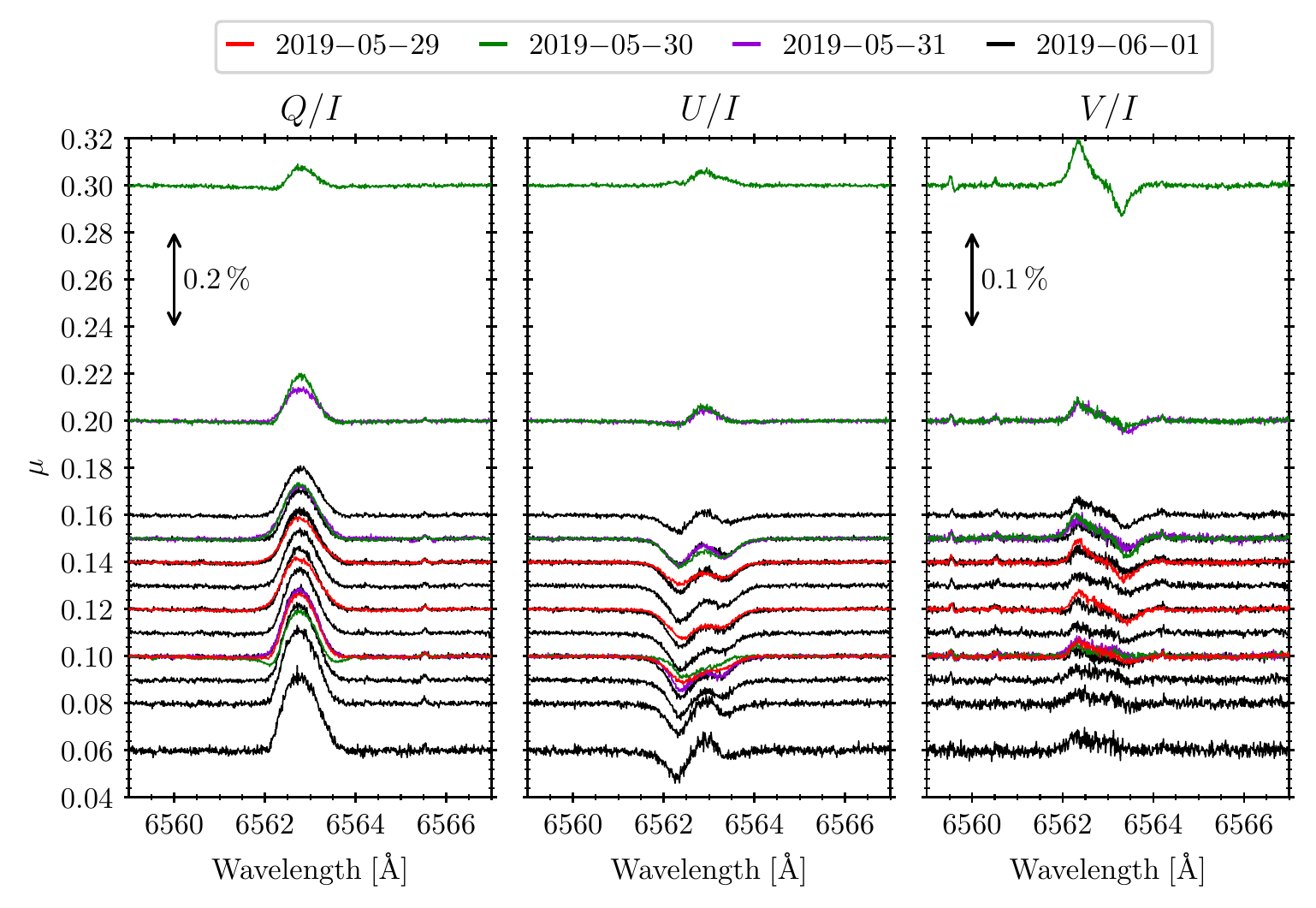}
    \caption{}
    \label{fig:clv_map_north}
  \end{subfigure}
  \begin{subfigure}[b]{0.82\textwidth}
    \includegraphics[width=\columnwidth]{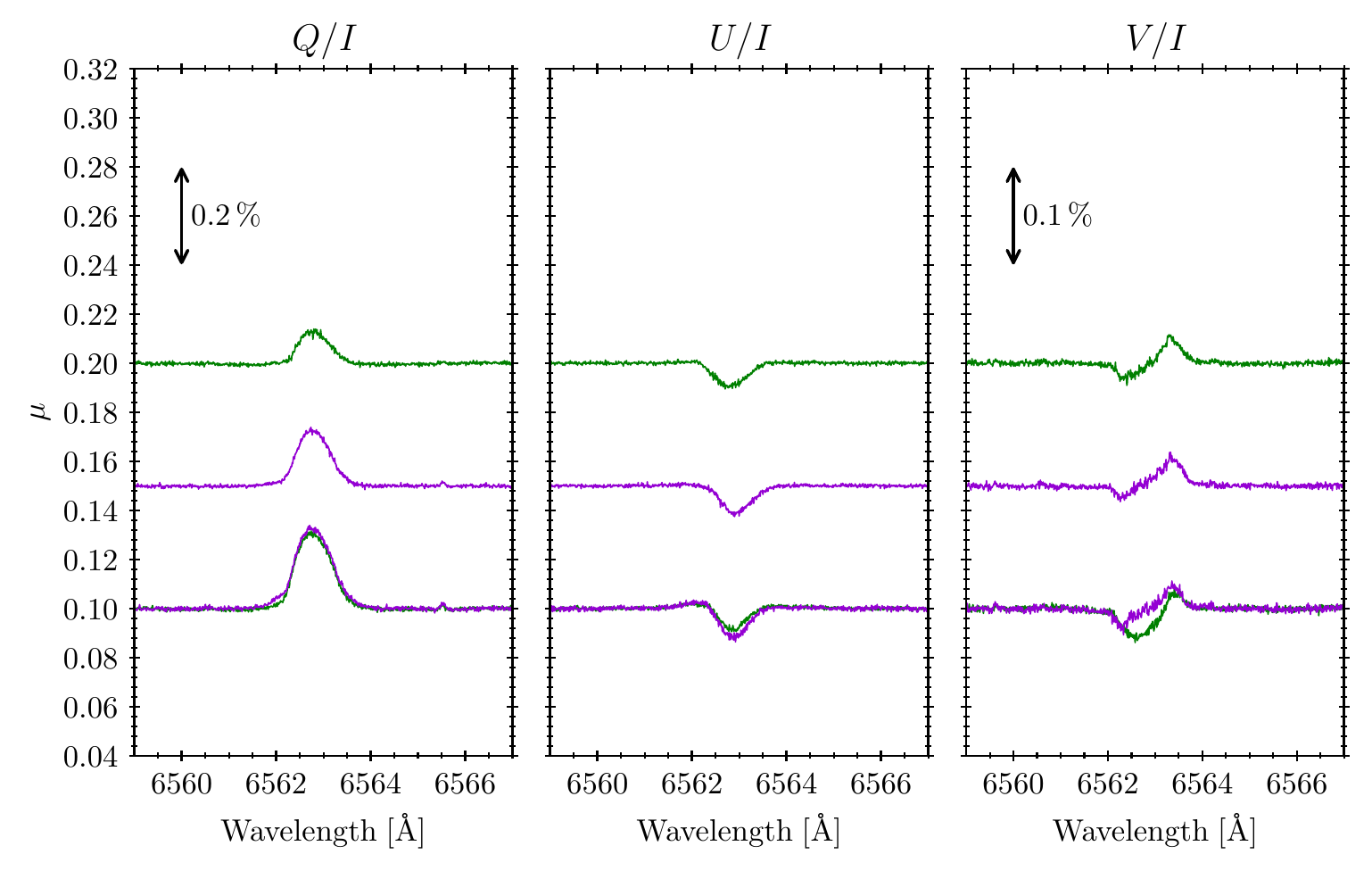}
    \caption{}
    \label{fig:clv_map_south}
  \end{subfigure}
  \caption{Center-to-limb variation of the $Q/I$, $U/I$, and $V/I$
    profiles of the emergent \ha radiation in coronal holes after
    spatial averaging. Panels (a) and (b) show observations at the
    North and South limbs, respectively. The vertical axis of the
    figures shows the $\mu$ position. The vertical row in the $Q/I$
    panels indicates the scale of the polarization amplitudes. The
    noise level, after averaging along the spatial direction of the
    slit, is about $3\times10^{-5}$. The reference direction for
    positive $Q/I$ is the parallel to the nearest limb. The ID of
    the North limb observations are 01, 02, 03, 05, 06, 07, 08, 23,
    24, 25, 28, 29, 30, 31, 32, 33, 34, 35, 36, 37 and 38. The
    South-limb observations are 09, 10, 26 and 27 (see Table
    \ref{tab:observations}).}
  \label{fig:clv_map}
\end{figure*}

After averaging all the spatial pixels of the slit covering a region
of $0\farcs5\times200\arcsec$, which correspond to a surface of about
50\,Mm$^2$ at the disk center, we obtain a single Stokes profile with
a noise level lightly above $10^{-5}$. We expect non-zero $Q/I$
signals for an observation near the limb because of the $90^{\circ}$
scattering geometry, as long as the reference direction for $Q>0$ is
parallel to the limb. Since the Stokes parameter $U$ does not have any
preferred direction, we expect zero $U/I$ signals if the averaged area
is large enough to statistically represent the quiet solar
atmosphere. The presence of thermal, dynamic and magnetic
inhomogeneities will generate non-zero $U/I$ signals. In this
subsection we analyze the spatially averaged profiles with the goal of
extracting information about the averaged properties of the
atmosphere.

Due to the curvature of the solar limb and the $0.5\arcsec$ precision
of the limb tracking system, some spatial pixels at the edges of the
slit can mix on-disk and off-limb signals, especially for observations
very close to the limb. Then, in order to consider spatial pixels
fully inside the disk, we select only the pixels that are at least
$1\farcs5$ inside the solar disk, which corresponds to pixels at
$\mu>0.05$, with $\mu=\cos\theta$ and $\theta$ the heliocentric
angle. After removing those pixels, we average the remaining pixels
and we obtain one profile for each Stokes parameter at any given limb
distance $\mu$.

Firstly, we analyze the CLV of the linear polarization observed in
coronal holes. We took several measurements with the slit parallel to
the limb at the North and South poles covering limb distances from
$\mu=0.06$ up to $\mu=0.3$. Figure \ref{fig:clv_map} shows the
averaged Stokes profiles at different limb distances, from the limb
(bottom profile) to the disk center (top profile). Each profile is
placed at a position on the vertical axis that corresponds to its
$\mu$ value. The different line colors correspond to different
observation days (see Table \ref{tab:observations}). Both, at the
North and South limbs, the $Q/I$ profiles are Gaussian-like and their
amplitudes decrease with the limb distance. The $U/I$ profiles at the
North pole are different from the ones observed at the South pole. The
later ones have negative Gaussian-like shapes with similar amplitudes
at different $\mu$ values. The profiles at the North limb show two
negative lobes, with the blue one larger than the red one, except the
ones at $\mu=0.06$ and $0.08$, which present a central positive
signal. The three-lobed $U/I$ profiles at the North limb show strong
positive bumps at the line-center with negative wings (see \fig
\ref{fig:maps2d_app_north}). These localized positive bumps are
detected in other North-limb observations from different days but not
in the observations at the South limb (see \ref{fig:maps2d_app_id09}).
Images from the Solar Dynamics Observatory (SDO) do not show any
appreciable differences on magnetic activity between both poles,
suggesting that these positive bumps in $U/I$ are probably due to
localized thermal or dynamic atmospheric conditions. The $V/I$ signals
are of the order of $0.02\,\%$ because there is no strong magnetic
activity, except at $\mu=0.3$ where the signal is twice as large.

\begin{figure} \centering
\includegraphics[width=1\columnwidth]{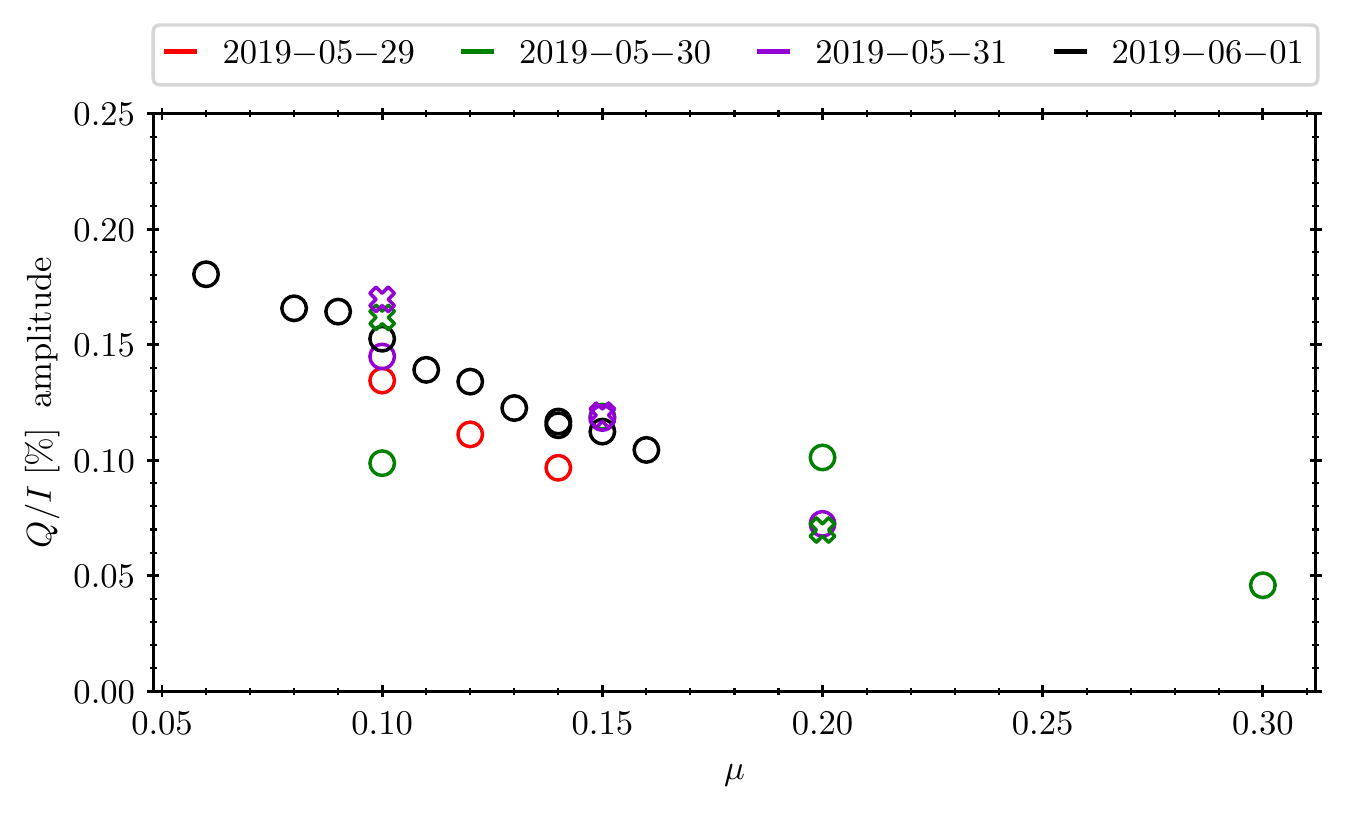}
\caption{Center-to-limb variation of the $Q/I$ amplitudes of the
  profiles of \fig \ref{fig:clv_map}. Observations taken at the North
  and South limbs are indicated by circles and crosses, respectively.}
  \label{fig:clv_holes}
\end{figure}

\fig \ref{fig:clv_holes} shows the CLV of the $Q/I$ amplitudes of the
profiles in \fig \ref{fig:clv_map}. The North and South limb
observations are represented by circles and crosses, respectively. As
we saw in \fig \ref{fig:clv_map}, the amplitudes decrease as we
approach the disk center. \fig \ref{fig:clv_holes} shows clearly that
the amplitudes of the measurements taken on \mbox{2019-05-29} (red
points, corresponding to the red lines in \fig \ref{fig:clv_map}) are
lower than those from \mbox{2019-06-01} (black points, corresponding
to the black lines in \fig \ref{fig:clv_map}). \fig \ref{fig:clv_map}
shows that the green profile at $\mu=0.1$ has a different shape, with
negative wings and a lower line center signal. And in this figure we
can clearly see that the amplitude has been reduced by 30\% with
respect to that of black points. This can be a manifestation of the
Hanle effect in the \ha line \cite[e.g.,][]{2011ASPC..437..117S,
  2011ApJ...732...80S} due to the contribution of the seven fine
structure components (see below).

\begin{figure*}
  \centering
  \includegraphics[width=1\textwidth]{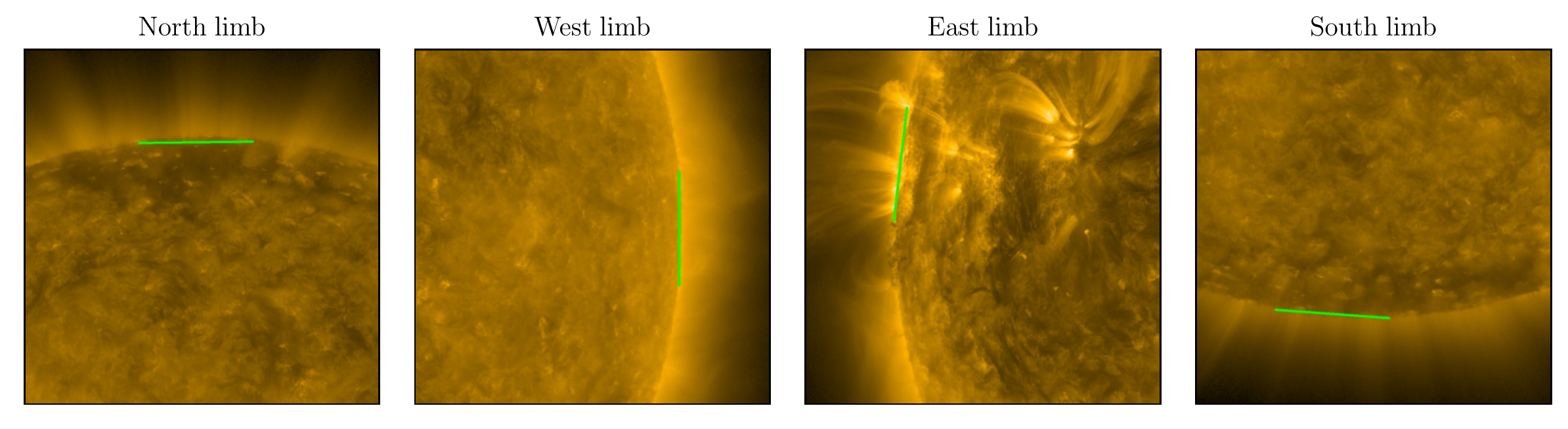}
  \caption{171 \r{A} SDO intensity images showing the location of the
    slit of our spectropolarimetric observations (see the green
    line). The ID of the observations are 01, 09, 12 and 40.}
  \label{fig:sdo_im}
\end{figure*}

\begin{figure*} \centering
\includegraphics[trim=0 190 0 0,clip,width=1\textwidth]{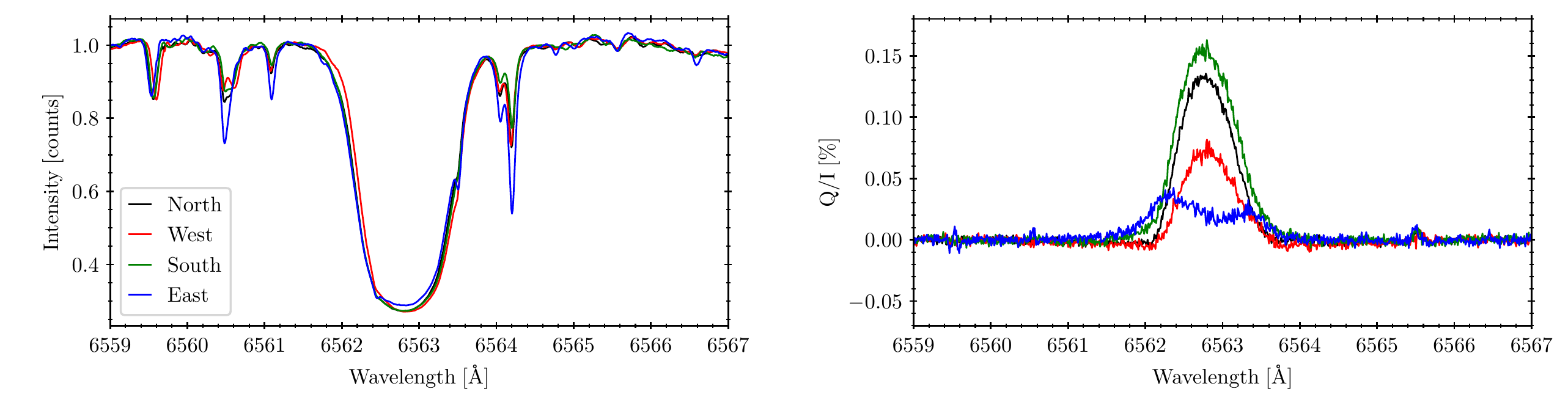}
\caption{Spatially averaged on-disk observations at $5\arcsec$ inside
  the limb ($\mu=0.1$) taken in different heliographic limbs. The ID
  of the observations are: 01, 09, 12 and 40 (see Table
  \ref{tab:observations}).}
  \label{fig:spat_avg}
\end{figure*}

It is also interesting to compare spatially averaged $Q/I$ profiles at
different solar regions. We consider four observations at $5\arcsec$
($\mu=0.1$) inside the solar disk. Two observations in the coronal
holes at the North and South limbs, one observation at the West limb
in a quiet region, and another one in a plage at the East limb. \fig
\ref{fig:sdo_im} shows the location of the slit for each observation
using SDO images at the 171\,\r{A} wavelength. The averaged Stokes
profiles at these locations are shown in \fig \ref{fig:spat_avg}. The
$Q/I$ profiles at the North and South limbs have Gaussian-like shapes
with similar amplitudes. At the West limb, the amplitude is twice
lower than at the North and South limbs, but the shape remains
Gaussian-like.  However, the shape of the $Q/I$ profile from the more
active East limb is a two-peaked profile, which is similar to the one
observed by \cite{2000sss..book.....G} but with a more pronounced
asymmetry. The lower $Q/I$ amplitude from the quiet West-limb profile
can be understood in terms of depolarization by the Hanle effect. In
the active East limb, the magnetic field may have larger gradients and
may be structured at larger spatial scales than in quiet regions. This
may facilitate that the different sensitivity of the overlapping \ha
transitions to the Hanle effect produce asymmetries in the $Q/I$ core.
A physical mechanism leading to the creation of asymmetries in the
linear polarization profiles of \ha has been proposed by
\cite{2010ApJ...711L.133S}. For additional developments on the role of
magnetic field gradients on the shape of the \ha linear polarization
profiles, see \cite{2011ASPC..437..117S}. The spectropolarimetric
images of the East limb observation are shown in \fig
\ref{fig:maps2d_east}. The other ones can be found in \figs
\ref{fig:maps2d_app_id01}, \ref{fig:maps2d_app_id09}, and
\ref{fig:maps2d_app_id12} of Appendix \ref{app:figures}.

In addition to the \ha line, we have been able to detect faint linear
and circular polarization signals in other spectral lines: Ti {\sc ii}
at $6559.56$\,\r{A} and Ti {\sc i} at $6565.50$\,\r{A}. $Q/I$ and
$V/I$ signals in the Ti {\sc ii} line are only detected in the regions
with strong magnetic fields, where the linear polarization is induced
by the Zeeman effect (East limb). We have also detected linear
polarization signals in the Ti {\sc i} line due to scattering with
amplitudes of the order of $0.01\,$\%, even in regions with strong
circular polarization.


\subsection{Spatial variation and line shapes\label{subsec:variation}}

\begin{figure*} \centering
  \begin{subfigure}[b]{1\textwidth}
    \includegraphics[width=0.95\linewidth]{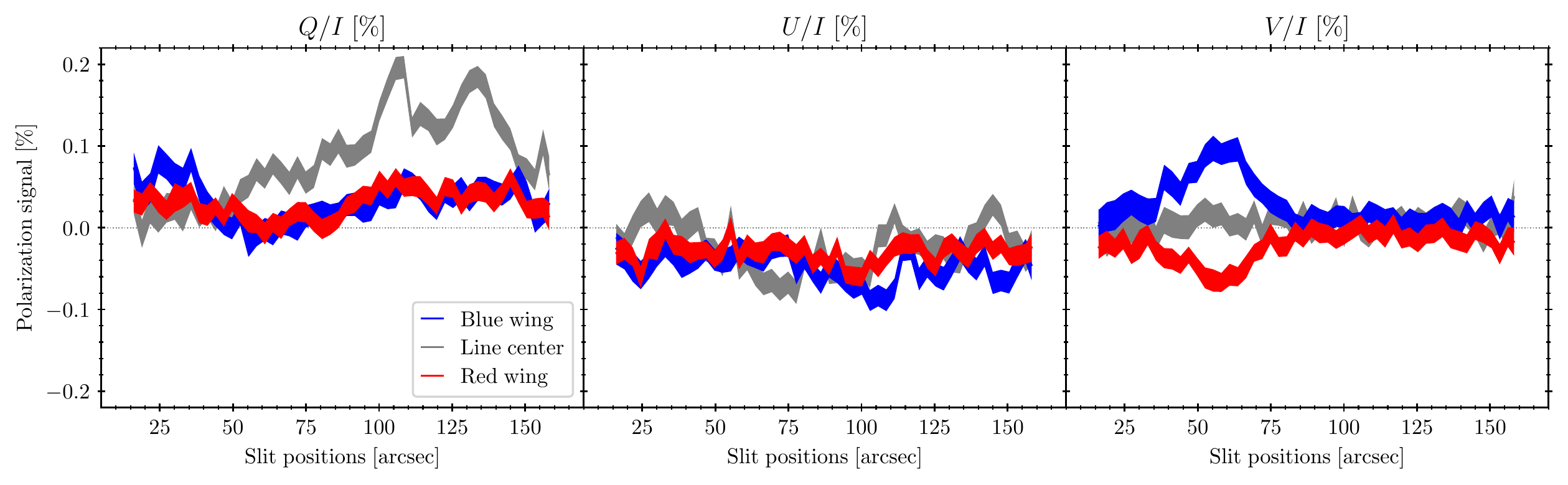}
    \caption{}
    \label{fig:variations_north}
  \end{subfigure}
  \begin{subfigure}[b]{1\textwidth} 
    \includegraphics[width=1\linewidth]{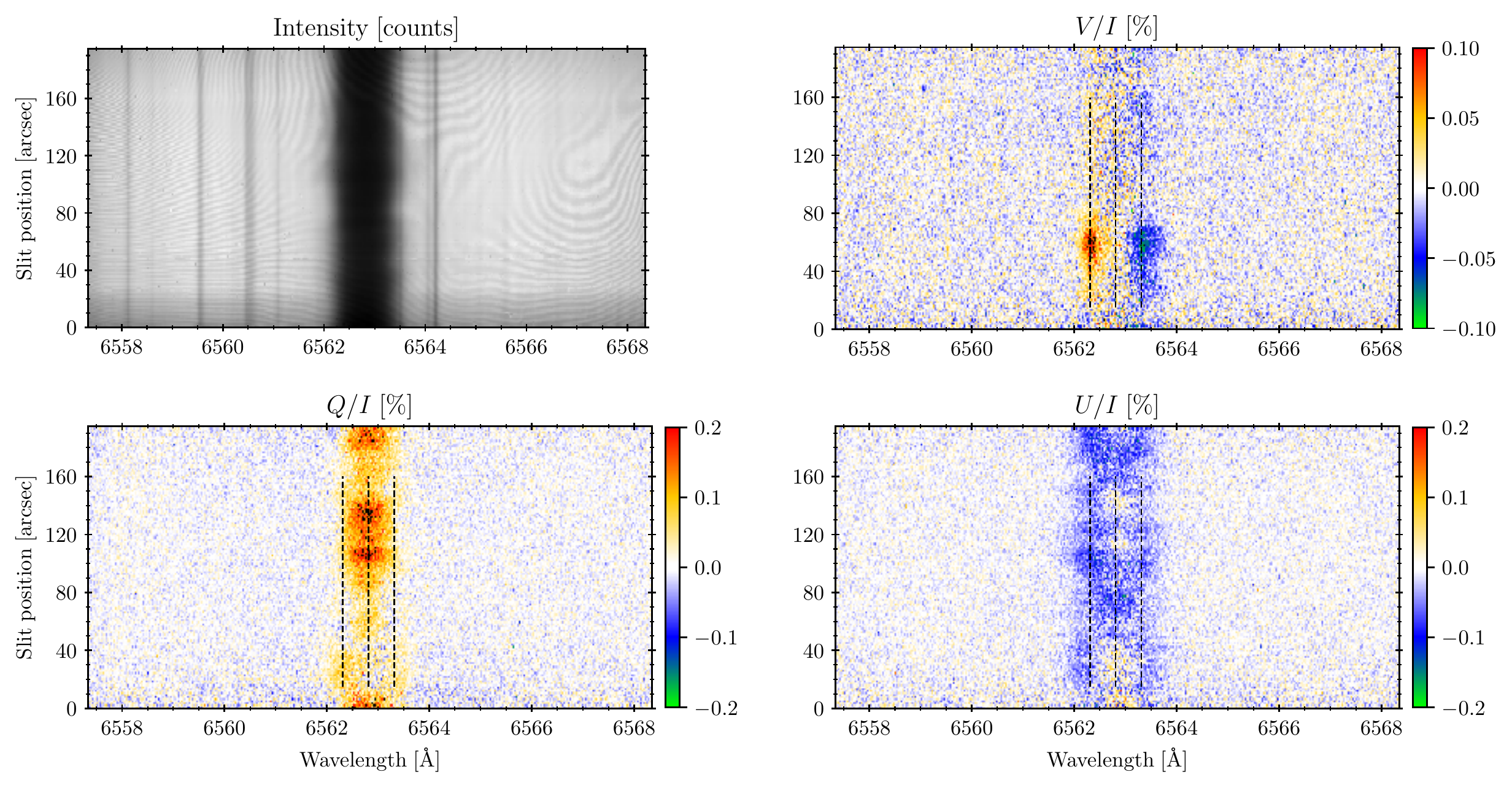}
    \caption{}
    \label{fig:maps2d_north}
  \end{subfigure}
  \caption{\textbf{Panel (a):} spatial variation at 3 different
    wavelengths along a region of the slit of $Q/I$ (left panel),
    $U/I$ (central panel) and $V/I$ (right panel). The line center
    wavelength (grey line) is at $\lambda_0=6562.81\,$\r{A}. The blue
    wing (blue line) is at $\lambda_b=\lambda_0-0.5\,$\r{A} and the
    red wing (red line) at $\lambda_r=\lambda_0+0.5\,$\r{A}. The width
    of the curves indicates the noise level at each spatial pixel. The
    considered spatial region is shown by the three dashed lines in
    the panels below. A binning of 2 and 4 pixels along the spatial
    and spectral dimensions are applied in order to reduce the
    noise. The spatial resolution of the signals are about
    $6\arcsec$. \textbf{Panel (b):} spectropolarimetric images. The
    color scale saturates in black at $\pm 0.2~\%$ for $Q/I$, and
    $U/I$ and $\pm 0.1~\%$ for $V/I$. Observation taken at the
    heliospheric North limb, at $\mu=0.14$. The observation ID is 03.}
  \label{fig:id03}
\end{figure*}

In the previous subsection, we showed how the amplitudes and shapes of
the linear polarization spatially averaged profiles change with the
limb distance. Such variations allow us to learn about the large-scale
structure of the chromosphere at different heights, since the
scattering signals of the \ha line originate at chromospheric heights
\mbox{\cite[][]{2011ApJ...732...80S}}. Clearly, small-scale variations
are also important in order to decipher how the magnetic and velocity
fields change locally.

As we specified in \sect \ref{ssec:instr}, the length of the slit is
$200\arcsec$ with a spatial sampling of $1.4\arcsec$. Since the seeing
of our observations is not worse than $2\arcsec$, we are limited by
the resolution of the CCD, which is twice the spatial sampling
(i.e. $\sim3\arcsec$). This allows us to see relatively small
variations on the Stokes profiles along the slit at different regions
of the solar disk at the expense of a lower signal-to-noise ratio
(SNR). Furthermore, it is important to remember that the integration
time of our observations is about 5 to 8 minutes, so that plasma
variations can smear or even cancel out the polarization signals
\cite[][and more references therein]{2017ApJ...843...64C}.

In \fig \ref{fig:variations_north}, we show the variation of the
polarization signals along the slit from an observation performed in a
coronal hole at the North limb, at $\mu=0.14$ with the slit parallel
to the limb at three different wavelengths: the line center
$\lambda_0$ and the near wings at
$\lambda_{r,b}=\lambda_0\pm0.5\,$\r{A}. The signals are binned over 2
and 4 spatial and spectral pixels, respectively. Therefore, the
spatial resolution is about $6\arcsec$. Figure \ref{fig:maps2d_north}
shows the spectropolarimetric images of the same observation to put in
context the other figure.

In \fig \ref{fig:variations_north} we can observe several
interesting things:

a) The wing signals of the $Q/I$ profiles decrease from $35\arcsec$ to
$60\arcsec$, while the $V/I$ amplitudes increase in this region. After
this position, the $V/I$ signal decreases until reaching $85\arcsec$
while the $Q/I$ signal wings increase until the $100\arcsec$ slit
position. Apparently, the $Q/I$ depolarization in the wings is related
to the magnetic fields.

b) The line-core signal of $Q/I$ increases almost linearly from
$45\arcsec$ up to $90\arcsec$ along the slit and the $U/I$ signal at
the line-center decreases, while the circular polarization signals
increase reaching a maximum of $0.1\,\%$ at $60\arcsec$ and then it
decreases. The linear polarization at the line core does not seem to
be affected by the magnetic fields in this region.

c) It seems that the signal in the wings of $U/I$ remains negative
without any significant change in the spatial region where $V/I$ is
detected.

d) At $25\arcsec$ slit position, the $Q/I$ signal in the blue wing is
twice the signal in the red wing, producing an asymmetric
profile. This asymmetry remains visible for $15\arcsec$ along the
slit. At these positions, the line-center signal of $U/I$ becomes
slightly positive while the wing signals stay negative. At $25\arcsec$
slit position, the $V/I$ signal starts increasing. The asymmetry
observed in $Q/I$ is similar to the observed by
\cite{2000sss..book.....G}, although in that case the signal was
integrated along a significantly larger spatial region.

e) From $90\arcsec$ slit position to the end, there is no sizable
circular polarization. However, we detect two maxima in $Q/I$ at
$105\arcsec$ and $130\arcsec$. Near these slit locations, the
line-center signal of $U/I$ becomes positive while the blue wing
signal slightly decreases. The dynamics of the plasma together with
the magnetic field, is capable of modifying the linear polarization
signals in a significant way \cite[e.g.,][]{2021ApJ...909..183J}.

\begin{figure*}\centering
  \includegraphics[width=1\linewidth]{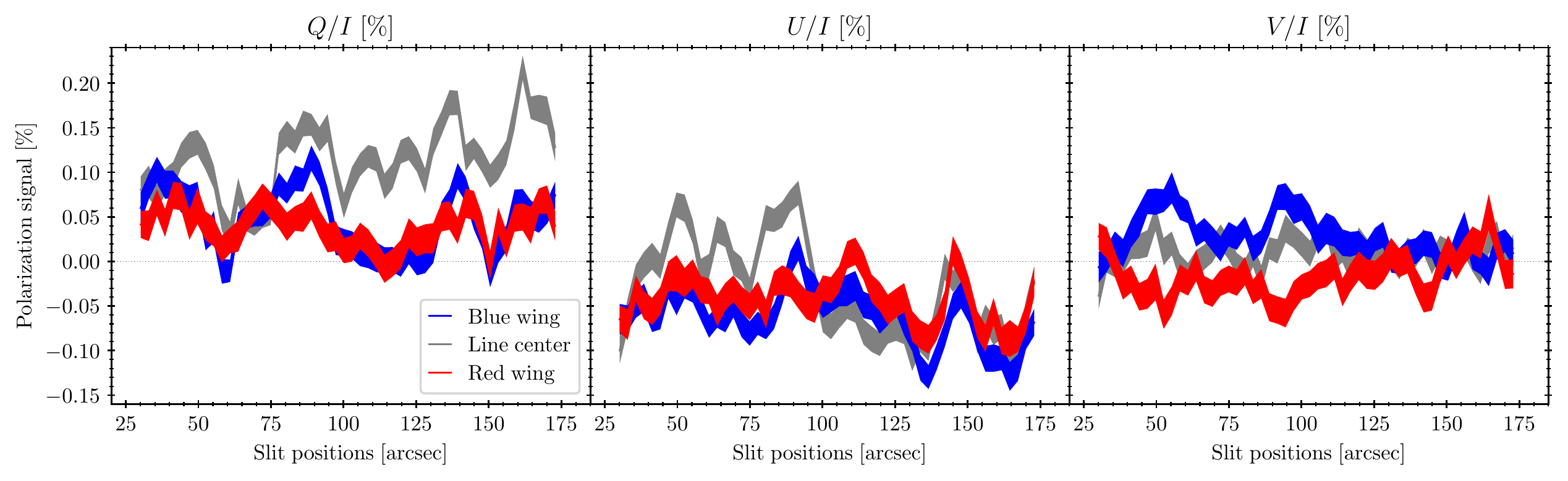}
  \caption{Similar figure to \fig \ref{fig:variations_north}, but for
    another observation at the North limb with ID 06. The
    spectropolarimetric image of this observation can be found in \fig
    \ref{fig:maps2d_app_north}.}
  \label{fig:variations_north_id06}
\end{figure*}

Figure \ref{fig:variations_north_id06} shows the spatial variation of
another North-limb observation at $\mu=0.15$, taken during a different
day. The $U/I$ signals have 3 positive peaks at the core, at the
$50\arcsec$, $65\arcsec$ and $90\arcsec$ positions, keeping a negative
signal in the wings. The line-center $Q/I$ signal also presents two
peaks at $50\arcsec$ and around $90\arcsec$. The $Q/I$ signals at the
wings also increase, but with a slightly larger amplitude in the blue
wing of the second peak. These spatial variations of the linear
polarization seem to be correlated with the signals detected in $V/I$,
reaching amplitudes of $\sim0.1\%$ at $50\arcsec$ and $95\arcsec$. On
the other hand, the $U/I$ signals and the wings of $Q/I$ suddenly
vanish at $\sim150\arcsec$. The $Q/I$ line-center signal does not
vanish at this position, but it decreases a factor two with respect to
previous and subsequent slit positions, reaching an amplitude of
$0.1\%$.

\begin{figure*} \centering
  \begin{subfigure}[b]{1\textwidth}
    \includegraphics[width=1\linewidth]{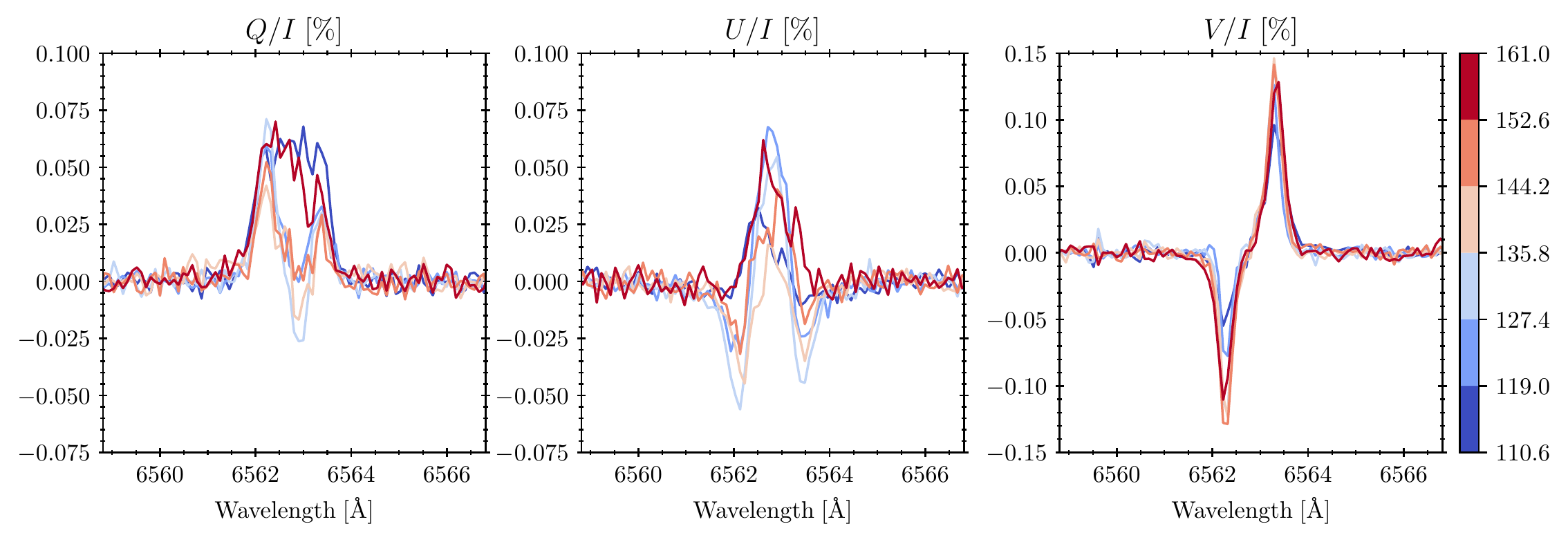}
    \caption{}
    \label{fig:var_east}
  \end{subfigure}
  \begin{subfigure}[b]{1\textwidth}
    \includegraphics[width=1\linewidth]{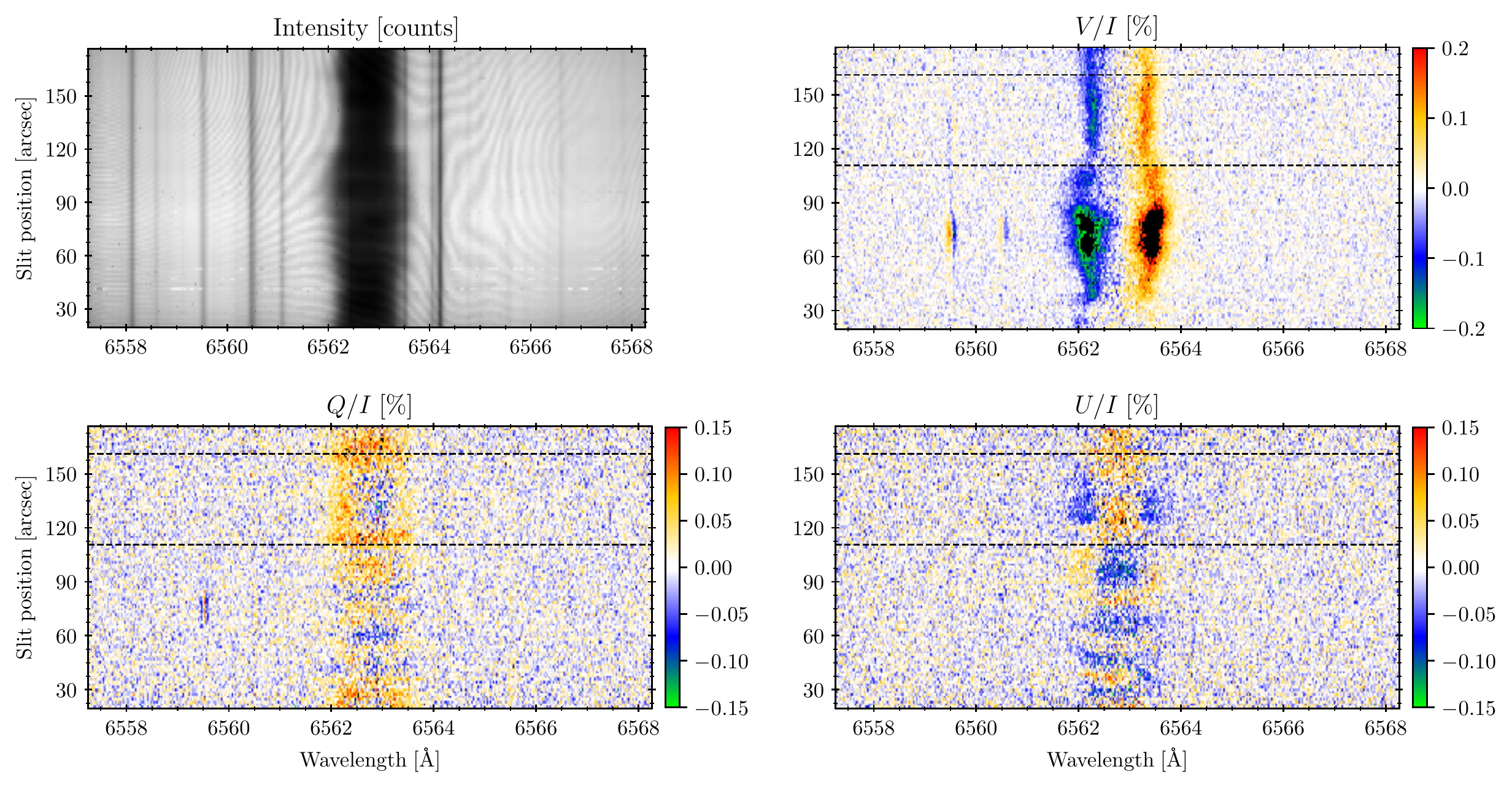}
    \caption{}
    \label{fig:maps2d_east}
  \end{subfigure}
  \caption{\textbf{Panel (a):} spatial variation of the \ha Stokes
    profiles shown in \fig \ref{fig:maps2d_east}. We applied a binning
    of 6 and 10 pixels along the spatial and spectral dimensions,
    respectively. Each profile, after the spatial binning, covers
    $8.4\arcsec$ on the slit. The selected region starts at
    $110.6\arcsec$ and ends at $161\arcsec$ (see the black thin lines
    in \fig \ref{fig:maps2d_east}). The ID of the observation used is
    40. \textbf{Panel (b):} spectropolarimetric images. The color
    scale saturates in black at $\pm 0.2~\%$ for $Q/I$, and $U/I$ and
    $\pm 0.1~\%$ for $V/I$. The black thin lines show the spatial
    region used in \fig \ref{fig:var_east}. Observation taken at the
    heliospheric East limb, at $\mu=0.1$. The observation ID is 40.}
\end{figure*}

Figure \ref{fig:maps2d_east} shows similar images but for an
observation taken at the East limb. This figure shows the spatial
variation of the blue profile of \fig \ref{fig:spat_avg}. This
observation shows large $V/I$ amplitudes with a maximum at
$60\arcsec-80\arcsec$. Along this spatial region of the slit, the
linear polarization profiles change several times their sign. This
could be an indication of different orientations of the magnetic
field. Unfortunately, the spatial resolution is not sufficient for
reaching more solid conclusions. However, if we focus on the $Q/I$
signals located between $110\arcsec$ and $160\arcsec$, we see that
they become two-peaked profiles with a negative signal in the core. We
have an inverted situation in the $U/I$ signals: negative wings with
positive signals in the core. In \fig \ref{fig:var_east} we emphasize
the spatial variation within this small region. We have performed a
binning of 6 and 10 pixels in the spatial and spectral dimensions,
respectively, in order to increase the SNR. The spatial location along
the slit is indicated by the color of the profiles, from blue to
red. Note that the dark blue linear polarization profile is fully
positive and the line core signal is flat, while the $V/I$ profile has
the lowest amplitude. As we move along the slit, towards redder
profiles, the $Q/I$ line-center signal becomes negative and its
profile asymmetric, the wings of the $U/I$ profile become negative and
the amplitude of the circular polarization increases, having a maximum
at intermediate positions and then it decreases. Interestingly, the
reddest $Q/I$ and $U/I$ profiles are similar to the bluest one, while
the $V/I$ amplitude is almost twice larger. This spatial variation
suggest that the non-symmetric shape of the linear polarization
profiles is produced by the magnetic field through the Hanle
effect. Note that the averaged blue profile shown in \fig
\ref{fig:spat_avg} has this two-peaked shape due to the signals from
this region, since outside of it the $Q/I$ signal is practically
cancelled out. We want to emphasize that this pattern in the circular
and linear polarization signals is detected in other observations (see
\fig \ref{fig:id03}) and it is almost certainly due to the interplay
of the Zeeman and Hanle effects at the different heights of formation
of the spectral line wavelengths. While the Zeeman effect is mostly
affecting the $V/I$ profile in the deeper atmospheric regions
\cite[][]{2004ApJ...603L.129S}, the Hanle effect plays a dominant role
in the formation of the linear polarization signals in the upper
chromosphere \cite[see Fig.~4 of][]{2010MmSAI..81..810S}.

The observation shown in \fig \ref{fig:maps2d_app_east} was taken at
the same limb, but on 30-05-2019, when the plage region of the East
limb was not yet visible. The circular polarization signals indicate
the presence of magnetic fields with components parallel to the LOS
along the whole slit, without strong variations. On the other hand,
the linear polarization signals vary significantly. Note that the
wings of the $Q/I$ and $U/I$ profiles also change spatially, something
that the previous observations did not show. Furthermore, the sign of
both $Q/I$ and $U/I$ profiles seem to be reversed: positive core
signals and negative wings in the former, while showing negative core
signals and positive wings in $U/I$.

The previous images show the Stokes parameters of the \ha radiation
emerging near the limb, where we expect significant amplitudes due to
the scattering geometry. However, \fig \ref{fig:maps2d_app_west} shows
an observation taken at $\mu=0.5$ at the West limb (quiet region) with
relatively large signals, $\sim 0.1\,\%$, that show an interesting
spatial variation along the slit. In this case, the limited spatial
resolution and the noise of the signal make it very difficult, nearly
impossible, to identify any pattern. Moreover, at our spatial
resolution the scattering polarization signals near the disk center
are expected to be faint and the current telescopes need large
exposure times to gain sufficient SNR.

It is of interest to remark that the high polarimetric sensitivity
achieved with ZIMPOL allows us to detect spatial variation of the
polarization signals of the Ti {\sc ii} line at
$\lambda=6559.56\,$\r{A} in strongly magnetized regions. In \fig
\ref{fig:maps2d_east}, we detect conspicuous Zeeman signals in
$Q/I$ and $V/I$. Although the signals are weak, we can distinguish
that the circular polarization profiles change their sign at about
$100\arcsec$. This means that $B_{||}$ at the formation height of this
line changes the sign at this location. Interestingly, the $V/I$
signal of \ha does not show any sign change there.

\subsection{Net circular polarization\label{subsec:net}}

\begin{figure*} \centering
  \begin{subfigure}[b]{1\textwidth}
    \includegraphics[width=1\linewidth]{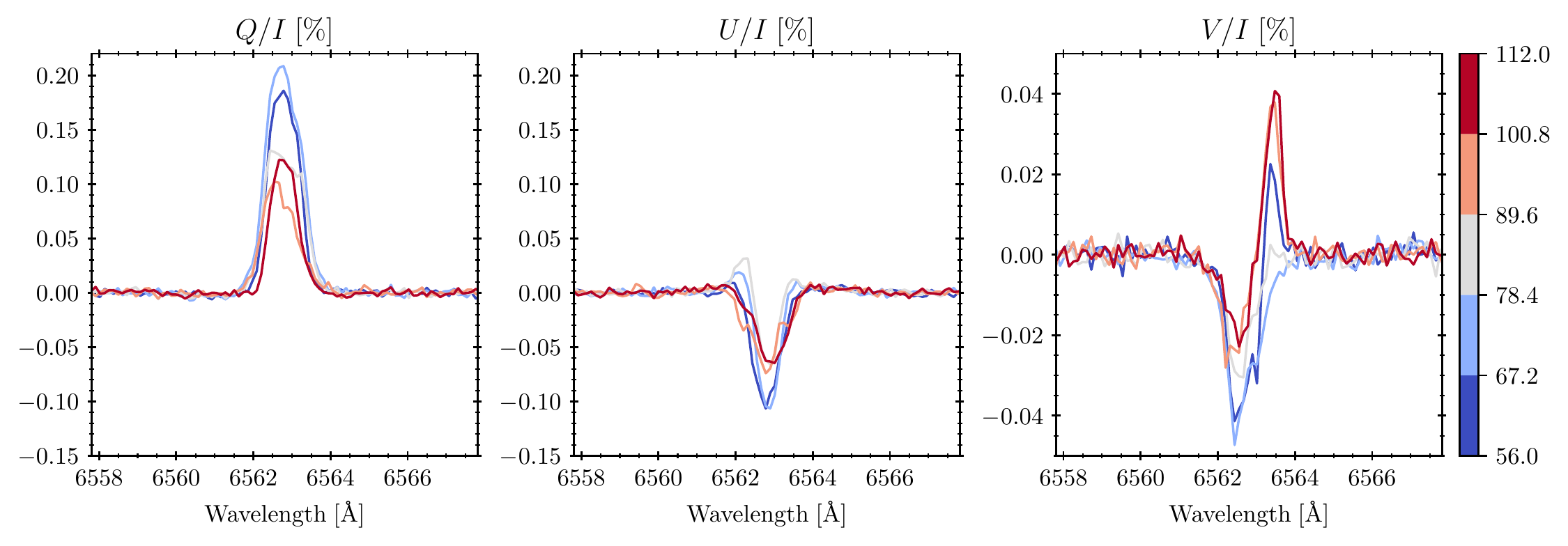}
    \caption{}
    \label{fig:ncp_spat_a}    
  \end{subfigure}
  \begin{subfigure}[b]{1\textwidth}
    \includegraphics[width=1\linewidth]{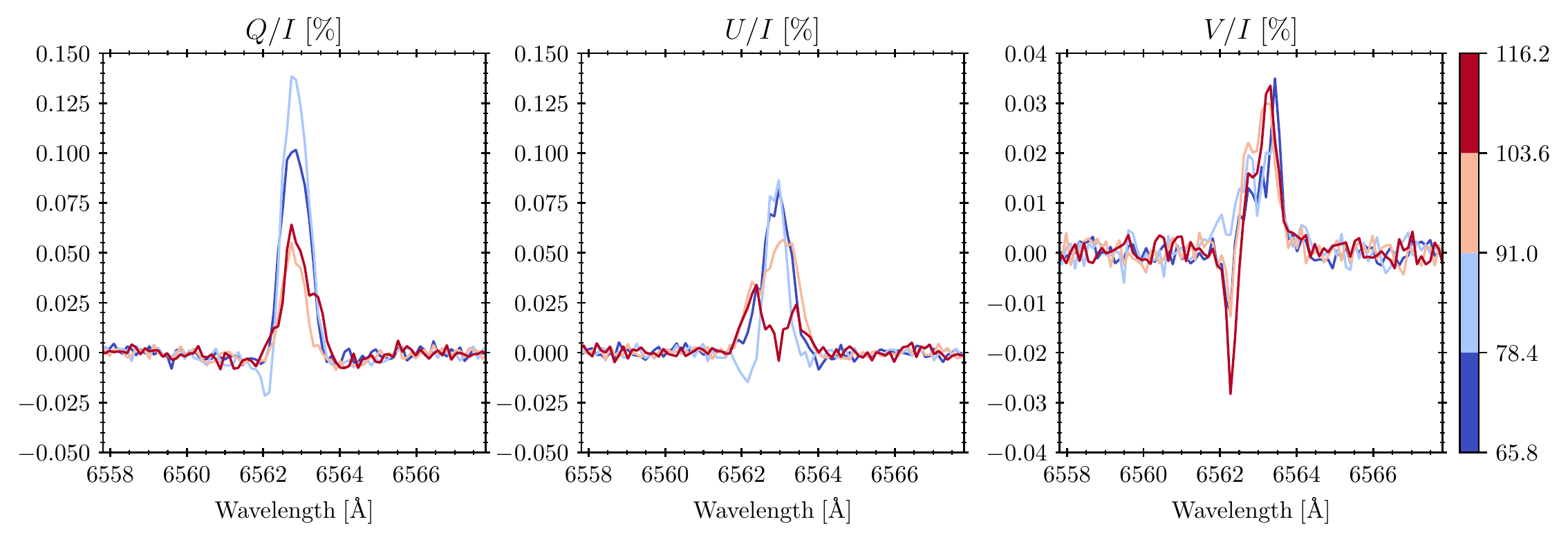}
    \caption{}
    \label{fig:ncp_spat_b}
  \end{subfigure}
  \caption{\textbf{Panel (a):} Stokes profiles at the center of the
    slit located at $\mu=0.1$. After averaging 8 spatial pixels, each
    profile covers $11.2\arcsec$ along the slit. The ID of the
    observation is 09. \textbf{Panel (b):} Stokes profiles at the
    center of the slit located at $\mu=0.09$. After averaging 9
    spatial pixels, each profile covers $12.6\arcsec$ along the
    slit. The ID of the observation is 11. Spatial variation of the
    Stokes profiles along the slit for different observations. A
    binning of 12 pixels in the spectral dimension is applied in both
    observations.}
  \label{fig:ncp_spat}
\end{figure*}

Net circular polarization (NCP) signals with $V/I$ amplitudes up to
$10^{-2}$ in the \ha line have been reported in prominences
\cite[see][]{2005ApJ...621L.145L}, although observations carried out
by \citet{2005ESASP.596E..82R} with a different spectropolarimeter
(ZIMPOL) could not confirm them. The latter observations could find
only NCP of a few times $10^{-4}$, which were considered to be
compatible with the alignment to orientation mechanism
\cite[][]{2004ASSL..307.....L}. Here we report NCP signals in quiet
regions near the limb. Observations with ZIMPOL taken near the solar
limb have to be considered carefully because the limb tracking
precision is limited and the resulting polarization profiles can mix
on-disk and off-limb solar disk signals. To avoid this problem, we
only selected slit pixels that are located at least $3\arcsec$ inside
the solar disk. With this condition, we can assure that the detected
polarization signals correspond to on-disk pixels, in spite of
possible deviations of the slit position due to occasional bad seeing
conditions and the limited limb tracking precision.

Figure \ref{fig:ncp_spat} shows two observations where NCP is
detected, as well as how the amplitudes and shapes change along the
slit. In order to improve the SNR, we average 12 spectral pixels, and
8 spatial pixels for the panel (a) and 9 spatial pixels for the panel
(b). Actually, the panel (a) shows the same spatially averaged
observation of \fig \ref{fig:spat_avg} (green curve; also shown in \fig
\ref{fig:maps2d_app_id09}). In panel (a) we see the reddest $V/I$
profile with an antisymmetric shape, but as we move along the slit (to
bluer profiles), the signal becomes fully negative. At the same time,
the linear polarization amplitudes increase as the shape of the
circular polarization $V/I$ profile becomes a one-lobe profile. At the
same time, the wings of the $U/I$ profiles become positive. We see a
similar behaviour in panel (b), but with the circular polarization
signal fully positive (this observation is also shown in \fig
\ref{fig:maps2d_app_id11}).

\subsection{The Weak Field Approximation\label{subsec:wfa}}

The \ha line is very broad because the hydrogen is the lightest atom
and its Doppler width ($\Delta\lambda_{\mathrm{D}}$) is usually much
larger than the Zeeman splitting of the levels
($\Delta\lambda_B$). That is one of the necessary conditions for
applying the so-called weak field approximation (WFA). The other
condition of the WFA validity is that the LOS component of the
magnetic field, $B_{||}$, is constant with optical depth. In fact,
this is a very strong assumption in the case of the \ha line since its
formation height spans few thousands of kilometers from the
photosphere to the top of the chromosphere. However, the method can
provide at least a rough estimate of the magnetic field strength with
very little effort. Therefore, we dedicate this section to such
quantitative analysis.

A detailed derivation of this approximation can be found in
\cite{2004ASSL..307.....L}. Under the assumptions mentioned above, the
WFA leads to an expression for the Stokes $V$ profile of the form
\begin{equation}
  \label{eq:wfa}
  V(\lambda) = - 4.6682\times 10^{-13}\,\lambda_0^2\,
  \bar{g}_{\mathrm{eff}}\, B_{||}\, \frac{\partial I(\lambda)}{\partial \lambda}\,,
\end{equation}
where $\lambda_0$ is in \r{A}, $B_{||}$ is the magnetic field
component along the LOS and it is in G, and $\bar{g}_{\mathrm{eff}}$
is the effective Land\'e factor of the \ha line which takes the value
1.048 following \cite{1994A&A...291..668C}. The investigation done by
\cite{2004ApJ...603L.129S} calculated the response functions in 1D
atmospheric models at $\mu=1$ and found that the $V$ signals are
mainly sensitive to photospheric magnetic fields. However,
\mbox{\cite{2012ApJ...749..136L}} showed that 1D radiative transfer
calculations are not reliable for synthesizing the emergent intensity
of this line, and that a 3D treatment is needed to model chromospheric
structures (e.g., fibril-like structures) using line-core intensity
maps. \cite{2020JApA...41...10N} applied the WFA to sunspots
spectropolarimetric observations in the \ha line. It is important to
remark that the \ha line is the result of seven blended radiative
transitions and the applicability of the WFA needs to be carefully
investigated. The fact that the synthesized line-core intensity in 3D
models traces chromospheric structures does not imply that the
retrieved $B_{||}$ through the WFA corresponds to chromospheric
heights. In a future theoretical investigation we want to evaluate the
reliability of the inferred magnetic fields through the WFA by means
of comparing the observations shown here and a detailed 3D radiative
transfer modelling. As a first step in this first paper, we apply the
WFA to infer and report the estimated magnetic field component along
the LOS in different regions of the solar disk with different magnetic
activities.

\begin{figure*} \centering
  \begin{subfigure}[b]{1\columnwidth}
    \includegraphics[width=0.95\linewidth]{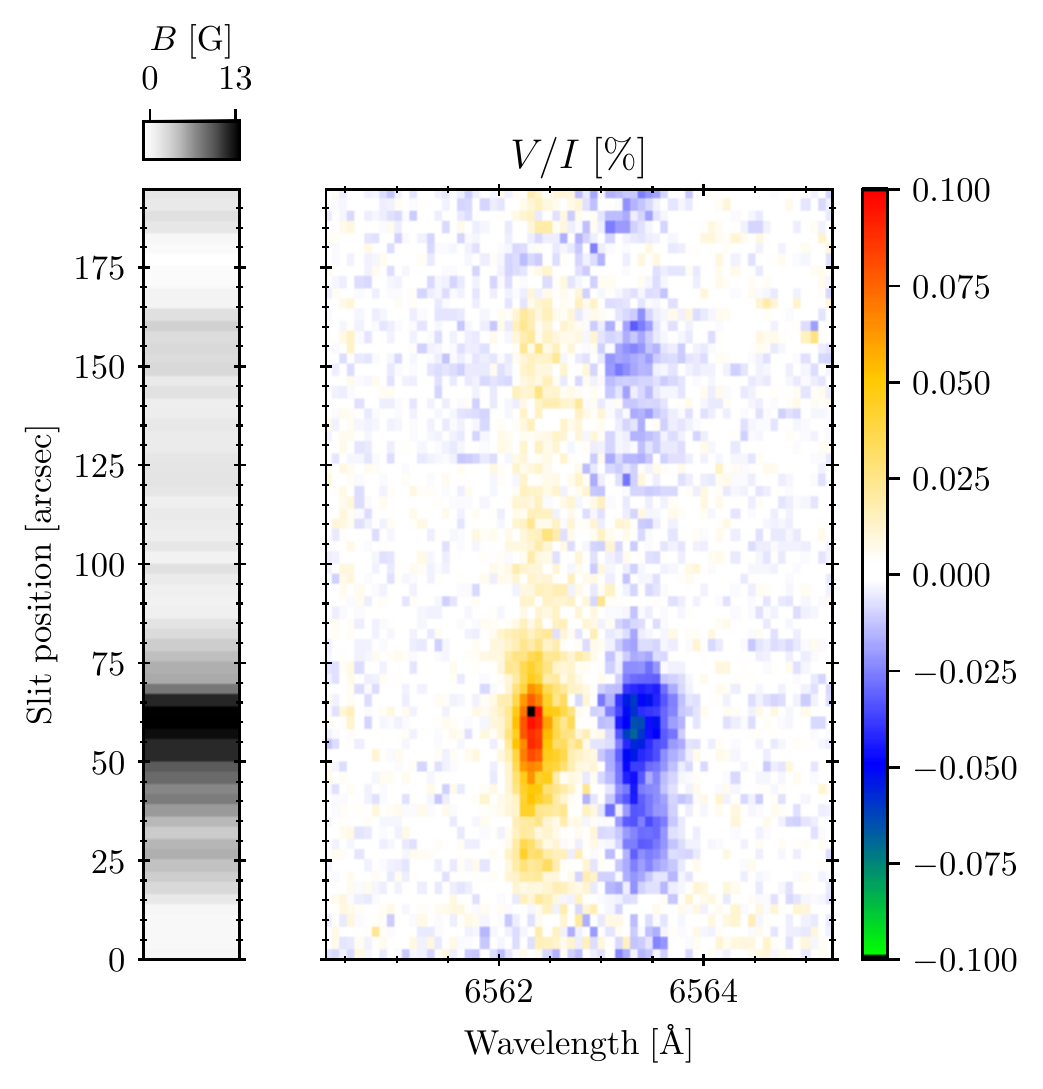}
    \caption{}
  \end{subfigure}
  \begin{subfigure}[b]{1\columnwidth}
    \includegraphics[width=0.95\linewidth]{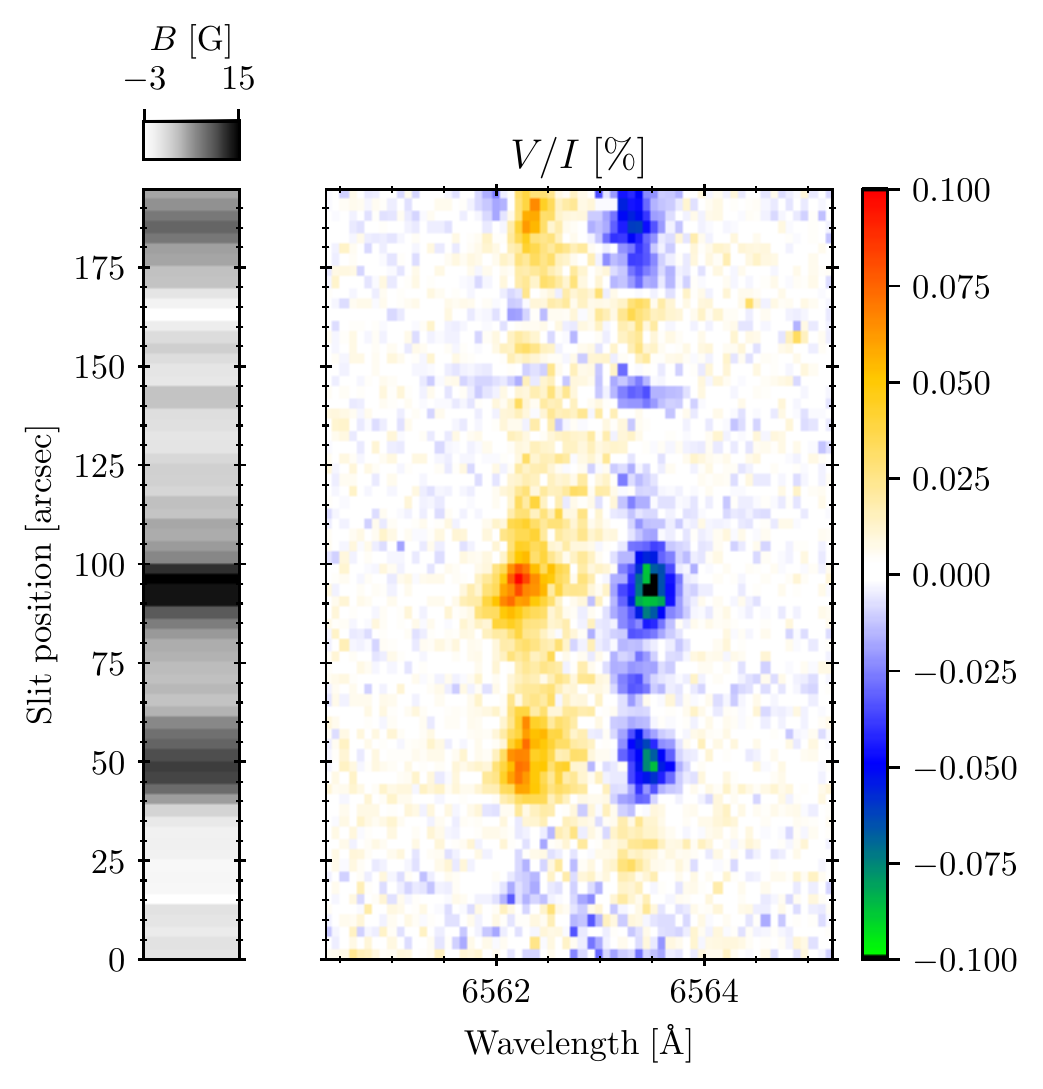}
    \caption{}
  \end{subfigure}
  \begin{subfigure}[b]{1\columnwidth}
    \includegraphics[width=0.95\linewidth]{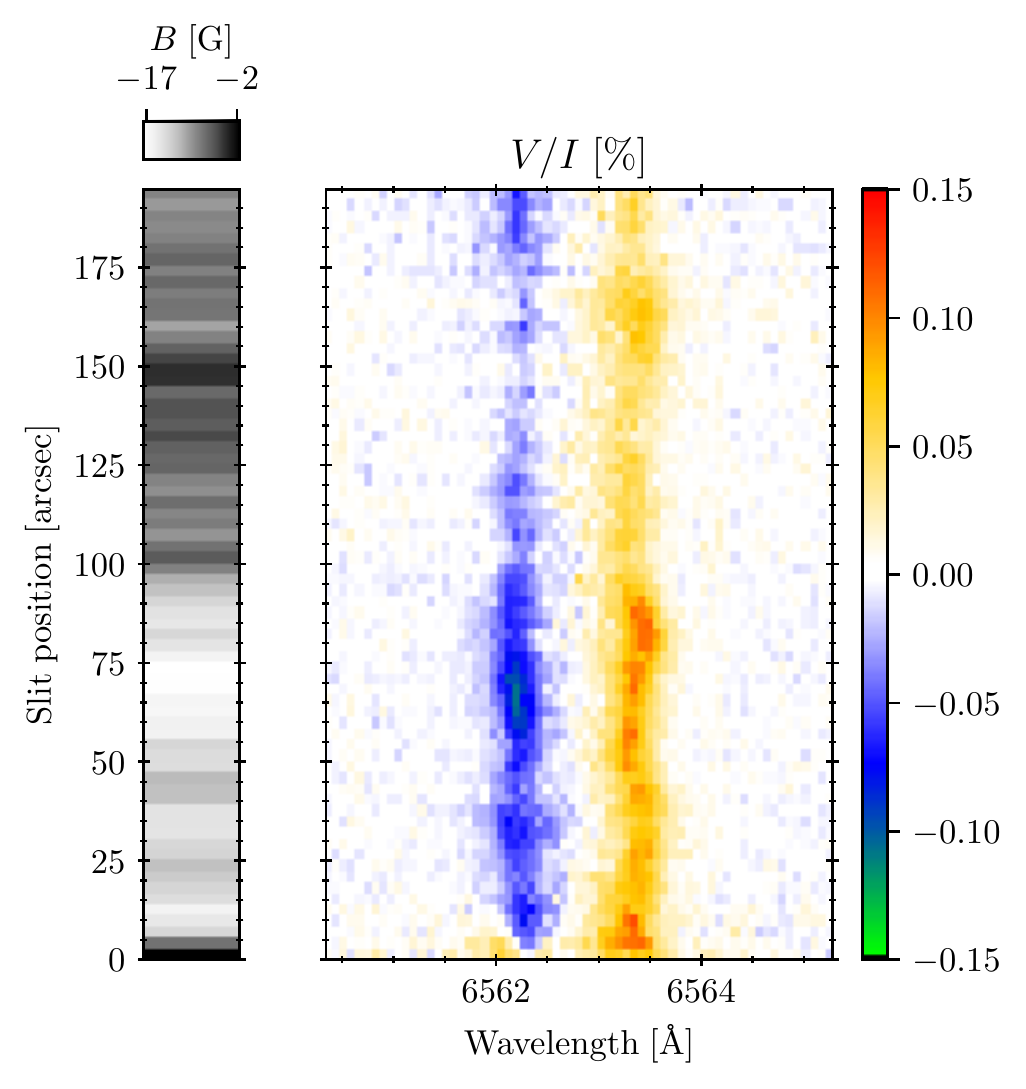}
    \caption{}
  \end{subfigure}
  \begin{subfigure}[b]{1\columnwidth}
    \includegraphics[width=0.95\linewidth]{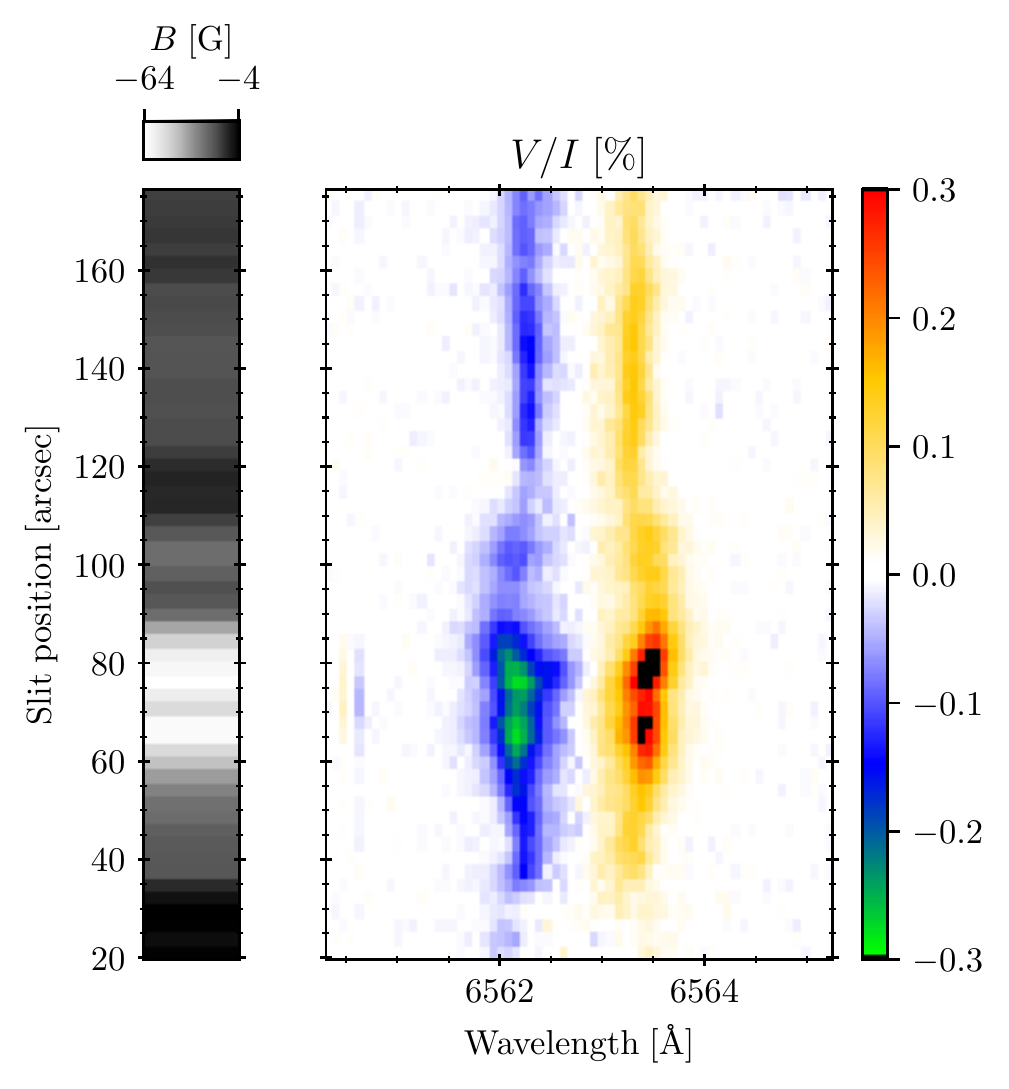}
    \caption{}
  \end{subfigure}
  \caption{\textbf{Panel (a):} slit located at North at
    $\mu=0.14$. Observation ID 03.. \textbf{Panel (b):} slit located
    at North at $\mu=0.15$. Observation ID 06. \textbf{Panel (c):}
    slit located at East at $\mu=0.13$. Observation ID
    22. \textbf{Panel (d):} slit located at East at $\mu=0.10$. ID
    40. Each panel is formed by two subpanels. \textbf{Left subpanel}:
    retrieved longitudinal component $B_{||}$ of the magnetic
    field. \textbf{Right subpanel}: the circular polarization
    signals.}
  \label{fig:wfa_halpha}
\end{figure*}

Assuming that the noise of our observations follows a Gaussian
distribution, we can apply the equations derived in
\cite{2012MNRAS.419..153M} in order to calculate $B_{||}$. To increase
the SNR, we applied a binning of 8 and 2 pixels in the spectral and
spatial dimensions, respectively. Moreover, we masked several spectral
lines near the \ha line that worsened the fit. Such lines and the
removed spectral region can be found in Table
\ref{tab:masked_lines}. We have to bear in mind that the spatial
resolution is low, and even more after binning. Then, the obtained
$B_{||}$ indicates the magnetic flux inside a spatial region of
approximately $3\arcsec\times 0.5\arcsec$.

\begin{table}
  \centering
  \caption{Removed spectral lines.}
  \label{tab:masked_lines}
  \begin{tabular}{ccl}
    $\lambda_0$~[\r{A}] & $\Delta\lambda$~[\r{A}] & Line \\
    \hline\hline
    6560.57 & 0.25 & Si {\sc i} \\
    6561.09 & 0.10 & Fe {\sc ii} \\
    6562.44 & 0.05 & V {\sc ii} \\
    6563.51 & 0.15 & Co {\sc i} \\
    6564.15  & 0.35 &  Unknown\\
    \hline
  \end{tabular}
  
  \begin{tablenotes}
    \small
  \item \textbf{Notes.} The first and second columns define the
    spectral region removed before applying the WFA:
    $\lambda_0\pm\Delta\lambda$. We note that $\Delta\lambda$ was set
    visually, checking the intensity profiles. The third column
    indicates the element of the transition in case it is known.
  \end{tablenotes}
\end{table}

If instead of retrieving the magnetic field for spatially averaged
profiles we consider the spatial variation of the measured Stokes
parameters, we can obtain $B_{||}$ along the slit. Figure
\ref{fig:wfa_halpha} shows the obtained $B_{||}$ at each spatial pixel
for four observations. The panel (a) and (b) (also shown in \fig
\ref{fig:maps2d_north} and \ref{fig:maps2d_app_north}) are
observations taken at the North limb. At the locations with larger
circular polarization signals, with amplitudes of $0.1\,$\%, we get
$B_{||}\approx15\,$G, while in regions where the signal is weak but
detectable we get magnetic fields of $2-3\,$G. The other two panels,
(c) and (d) (shown in \fig \ref{fig:maps2d_app_east} and
\ref{fig:maps2d_east}), are observations taken at the East limb. The
obtained magnetic field in panel (c) is more or less constant along
the slit, reaching a maximum of $18\,$G. On the other hand, we see
much stronger circular polarization signals in the panel (d) at some
locations, getting $B_{||}\approx 65\,$G. It is important to remember
that, as stated above, the obtained values are only estimations of the
actual magnetic field and it is not possible to know the error of
applying this approximation. The last panel shows an observation taken
in a plage region near the limb, and we have to bear in mind that the
retrieved magnetic field strengths correspond to the magnetic fields
parallel to our LOS. Then, if we make the assumption that the magnetic
field permeating the plage is vertical, the $B_{||}$ component is the
product of $B$ by $\mu$. Roughly speaking, the retrieved magnetic
field along the LOS is one order of magnitude lower than the true
magnetic field strength.


\section{Conclusions\label{sec:conclusions}}

Motivated by the need to develop new methods to probe the magnetism of
the solar chromosphere, we have initiated an observational and
theoretical research project aimed at clarifying the diagnostic
potential of spectropolarimetry in the hydrogen \ha line. In this
respect, the $Q/I$ and $U/I$ profiles of this spectral line are of
particular interest because they are dominated by scattering
processes, and via the Hanle effect the line-center signals are
sensitive to the presence of magnetic fields in the upper solar
chromosphere \citep{2010ApJ...711L.133S}. In contrast, the circular
polarization signals are dominated by the Zeeman effect, and they
probe deeper atmospheric layers.

In this first paper we have presented an overview of our
spectropolarimetric observations of close to the limb regions of the
solar disk outside sunspots. The observed regions have varying
degrees of magnetic activity, and from the application of the WFA to
the observed Stokes $I$ and $V$ signals we have estimated the averaged
longitudinal field strength over the spatio-temporal resolution
element of the observation.

Concerning the scattering polarization signals, we have given
particular attention to the CLV curves corresponding to each of the
observed regions, since they are of great interest for our future
confrontations with the results from radiative transfer modelling in 3D
models of the solar atmosphere. A curious finding is that the $U/I$
profiles from the South and North limbs are different, in spite of the
fact that our spatial averaging is smearing out all the local
variations of the signals, which suggests different physical
conditions in the respective coronal holes. The South limb $U/I$
profiles are fully negative, while the North limb ones have negative
wings with the line-core close to zero or even positive (see \fig
\ref{fig:clv_map}). However, the $Q/I$ profiles from both limbs are
practically identical. On the other hand, the amplitudes of the $Q/I$
profiles from both limbs decrease as we approach the disk center (see
\fig \ref{fig:clv_holes}), getting an amplitude of $0.18\,\%$ at
$\mu=0.06$ and of $0.05\,\%$ at $\mu=0.3$. The spatially averaged
$Q/I$ signals in \fig \ref{fig:spat_avg} show fingerprints of the
presence of magnetic fields via the Hanle effect. The blue curve
(corresponding to the more magnetic region, see \fig \ref{fig:sdo_im})
has lower amplitude than the others. Moreover, this profile loses the
Gaussian shape and becomes a two-peaked profile, being similar to the
signal observed by \cite{2000sss..book.....G}.

We have also analyzed in detail the spatial variation of the Stokes
profiles along the direction of the spectrograph's slit, finding a
rich variability in both the linear and circular polarization
signals. Such spatial variations are not easy to interpret because the
\ha line is a multiplet with several overlapped transitions that are
formed at slightly different heights in the inhomogeneous plasma of
the solar chromosphere.  However, we have been able to identify some
patterns relating the linear and circular polarization signals, which
may be useful for the interpretation of the observations. In most
cases, in the presence of significant $V/I$ signals the core of the
$Q/I$ profiles are reduced and become a two-peaked profile, or they
just loose their Gaussian shape and turn into an asymmetric
profile. Some examples can be seen in \fig \ref{fig:maps2d_north},
\ref{fig:maps2d_east} and
\ref{fig:maps2d_app_north}. \cite{2011ApJ...732...80S} performed a 1D
radiative transfer investigation of the generation and transfer of the
scattering polarization in \ha, and concluded that the presence of
spatial gradients in the strength of the magnetic field in the upper
chromosphere can produce a line-core asymmetry in the $Q/I$ profile,
such as that observed by \cite{2000sss..book.....G} without
spatio-temporal resolution. This possibility could explain the
deformation we have observed in the $Q/I$ profile.

Interestingly, in our observations we have detected net circular
polarization signals at some positions along the spatial direction of
the slit in chromospheric regions. Furthermore, we have been able to
detect spatial variations of the linear and circular polarization
signals near such locations. In two different observations at
different solar limbs, the linear polarization amplitude increases as
the circular polarization becomes a one-lobe profile. These variations
could put some constraints for deciphering the physical mechanism that
creates such NCP signals. The NCP detected in \fig
\ref{fig:maps2d_app_id09} at $75\arcsec$ is remarkable. \fig
\ref{fig:ncp_spat_a} shows how the linear polarization amplitude
increases as the circular polarization becomes negative. Similar
signals are found in \fig \ref{fig:ncp_spat_b}, which corresponds to
the spectropolarimetric images shown in
\ref{fig:maps2d_app_id11}. Although the Stokes V signals are still
noisy after averaging some pixels and some features cannot be fully
trusted, the NCP profiles lie above the noise level and we believe
they are real. Given that the densities and temperatures in
prominences are similar to those of the upper chromosphere, a
theoretical explanation based on the semi-classical impact
approximation theory of collisions of hydrogen, protons, and electrons
in the presence of magnetic fields could perhaps provide a theoretical
explanation of these curious NCP observations
\citep{2008sf2a.conf..573S}. If such a hypothesis is confirmed by
close-coupling calculations, it could provide very strong constrains
on the plasma density and the magnetic field vector.

The high polarimetric sensitivity of our observations with ZIMPOL at
IRSOL has allowed us to detect an extremely rich spatial variability
of the scattering polarization of the \ha line, in spite of the low
spatial resolution of this instrumental setup.  This strongly
motivates us to continue spectropolarimetric observations of the \ha
line using the new generation of solar telescopes, namely the Daniel
K. Inouye Solar Telescope \cite[DKIST;][]{rimmele-2020} and the future
European Solar Telescope \cite[EST;][]{2013MmSAI..84..379C}. Equally
important is to theoretically investigate how are the Stokes profiles
of the \ha line in today's 3D models of the solar atmosphere, taking
into account the joint action of scattering processes and the Hanle
and Zeeman effects. This will be the topic of our next paper.

\begin{acknowledgements}
  J.J.B. acknowledges financial support from the Spanish Ministry of
  Economy and Competitiveness (MINECO) under the 2015 Severo Ochoa
  Programme MINECO SEV--2015--0548. J.T.B. acknowledges the funding
  received from the European Research Council (ERC) under the European
  Union's Horizon 2020 research and innovation program (ERC Advanced
  Grant agreement \mbox{No.~742265}), as well as through the projects
  \mbox{PGC2018-095832-B-I00} and \mbox{PGC2018-102108-B-I00} of the
  Spanish Ministry of Science, Innovation and
  Universities. J.\v{S}. acknowledges the financial support of grant
  \mbox{19-20632S} of the Czech Grant Foundation (GA\v{C}R) and from
  project \mbox{RVO:67985815} of the Astronomical Institute of the
  Czech Academy of Sciences. R.R. and M.B. acknowledge the support of
  the Swiss National Science Foundation (SNF) through grant
  200020-184952.  IRSOL is supported by the Swiss Confederation
  (SERI), Canton Ticino, the city of Locarno and the local
  municipalities.
\end{acknowledgements}

\bibliography{mybibtex}{} \bibliographystyle{aa}

\clearpage
\begin{appendix}
\onecolumn
\section{Table of observations}
\begin{table*}[h!]
  \centering
  \caption{Information about the observations.}
  \label{tab:observations}
  \begin{tabular}{lccccr}
    ID & Date (UTC) & Exposure time (s) & $\mu=\cos\theta$ & $\sigma_{\mathrm{noise}}[\times 10^{-4}]$ & Slit position \\
    \hline\hline
    01 & 2019-05-29 - 07:44:23 & 569 & 0.10 & $1.6$ & North \\
    02 & 2019-05-29 - 08:08:14 & 569 & 0.12 & $1.8$ & North \\
    03 & 2019-05-29 - 08:18:41 & 568 & 0.14 & $1.8$ & North \\
    04 & 2019-05-29 - 08:29:42 & 568 & 0.00 & $2.3$ & North \\
    05 & 2019-05-30 - 07:47:13 & 569 & 0.10 & $1.6$ & North \\
    06 & 2019-05-30 - 07:58:17 & 549 & 0.15 & $1.6$ & North \\
    07 & 2019-05-30 - 08:23:39 & 345 & 0.20 & $2.3$ & North \\
    08 & 2019-05-30 - 08:34:35 & 324 & 0.30 & $2.3$ & North \\
    09 & 2019-05-30 - 09:20:58 & 272 & 0.10 & $2.2$ & South \\
    10 & 2019-05-30 - 09:36:21 & 324 & 0.20 & $2.4$ & South \\
    11 & 2019-05-30 - 15:10:01 & 344 & 0.09 & $2.3$ & West \\
    12 & 2019-05-30 - 15:16:07 & 344 & 0.10 & $2.3$ & West \\
    13 & 2019-05-30 - 15:22:12 & 344 & 0.11 & $2.4$ & West \\
    14 & 2019-05-30 - 15:28:18 & 344 & 0.12 & $2.4$ & West \\
    15 & 2019-05-30 - 15:35:27 & 344 & 0.13 & $2.4$ & West \\
    16 & 2019-05-30 - 15:41:41 & 345 & 0.14 & $2.4$ & West \\
    17 & 2019-05-30 - 15:48:13 & 345 & 0.14 & $2.4$ & West \\
    18 & 2019-05-30 - 15:57:57 & 286 & 0.50 & $2.5$ & West \\
    19 & 2019-05-30 - 16:22:28 & 344 & 0.10 & $2.5$ & East \\
    20 & 2019-05-30 - 16:30:39 & 344 & 0.11 & $2.6$ & East \\
    21 & 2019-05-30 - 16:37:05 & 345 & 0.12 & $2.7$ & East \\
    22 & 2019-05-30 - 16:43:21 & 343 & 0.13 & $2.6$ & East \\
    23 & 2019-05-31 - 09:22:15 & 344 & 0.10 & $2.2$ & North \\
    24 & 2019-05-31 - 09:30:09 & 345 & 0.15 & $2.3$ & North \\
    25 & 2019-05-31 - 09:55:03 & 324 & 0.20 & $2.3$ & North \\
    26 & 2019-05-31 - 10:13:49 & 344 & 0.10 & $2.6$ & South \\
    27 & 2019-05-31 - 10:21:12 & 344 & 0.15 & $2.4$ & South \\
    28 & 2019-06-01 - 07:20:21 & 282 & 0.06 & $2.9$ & North \\
    29 & 2019-06-01 - 07:26:12 & 282 & 0.08 & $2.7$ & North \\
    30 & 2019-06-01 - 07:31:34 & 281 & 0.09 & $2.6$ & North \\
    31 & 2019-06-01 - 07:36:58 & 282 & 0.10 & $2.6$ & North \\
    32 & 2019-06-01 - 07:48:01 & 282 & 0.11 & $2.6$ & North \\
    33 & 2019-06-01 - 07:53:25 & 282 & 0.12 & $2.6$ & North \\
    34 & 2019-06-01 - 07:59:35 & 282 & 0.13 & $2.6$ & North \\
    35 & 2019-06-01 - 08:04:38 & 281 & 0.14 & $2.5$ & North \\
    36 & 2019-06-01 - 08:16:42 & 282 & 0.14 & $2.5$ & North \\
    37 & 2019-06-01 - 08:21:51 & 281 & 0.15 & $2.4$ & North \\
    38 & 2019-06-01 - 08:27:00 & 282 & 0.16 & $2.4$ & North \\
    39 & 2019-06-02 - 15:39:54 & 344 & 0.08 & $2.6$ & East (plage) \\
    40 & 2019-06-02 - 15:46:32 & 344 & 0.10 & $2.6$ & East (plage) \\
    41 & 2019-06-02 - 15:52:57 & 343 & 0.12 & $2.7$ & East (plage) \\
    42 & 2019-06-02 - 15:59:55 & 344 & 0.14 & $2.7$ & East (plage) \\
    43 & 2019-06-02 - 16:06:29 & 344 & 0.15 & $2.6$ & East (plage) \\
    44 & 2019-06-02 - 16:12:52 & 344 & 0.16 & $2.7$ & East (plage) \\
    45 & 2019-06-02 - 16:21:05 & 345 & 0.38 & $2.5$ & East (faculae) \\
    46 & 2019-06-02 - 16:29:40 & 285 & 0.08 & $3.4$ & West \\
    47 & 2019-06-02 - 16:37:40 & 284 & 0.10 & $3.3$ & West \\
    \hline
  \end{tabular}

  \begin{tablenotes}
    \small
  \item \textbf{Notes.} \textbf{First column:} identifier of the
    observation. \textbf{Second column:} date and time of the
    observation in UTC. \textbf{Third column:} exposure time in
    seconds for each of the two measurements of each run,
    $(I,\,Q/I,\,V/I)$ and $(I,\,U/I,\,V/I$). \textbf{Fourth column:}
    cosine of the heliocentric angle. \textbf{Fifth column:} standard
    deviations of the noise per pixel in the polarization
    images. \textbf{Last column:} slit position on the solar disk. The
    plage at the East limb was not visible until the 31th of May. For
    this reason, the observations of the 2nd of June has been
    differentiated from the ones on the 31th of May.
  \end{tablenotes}
\end{table*}

\clearpage
\section{Additional figures}
\label{app:figures}

\begin{figure*}[h!] \centering
\includegraphics[width=0.96\linewidth]{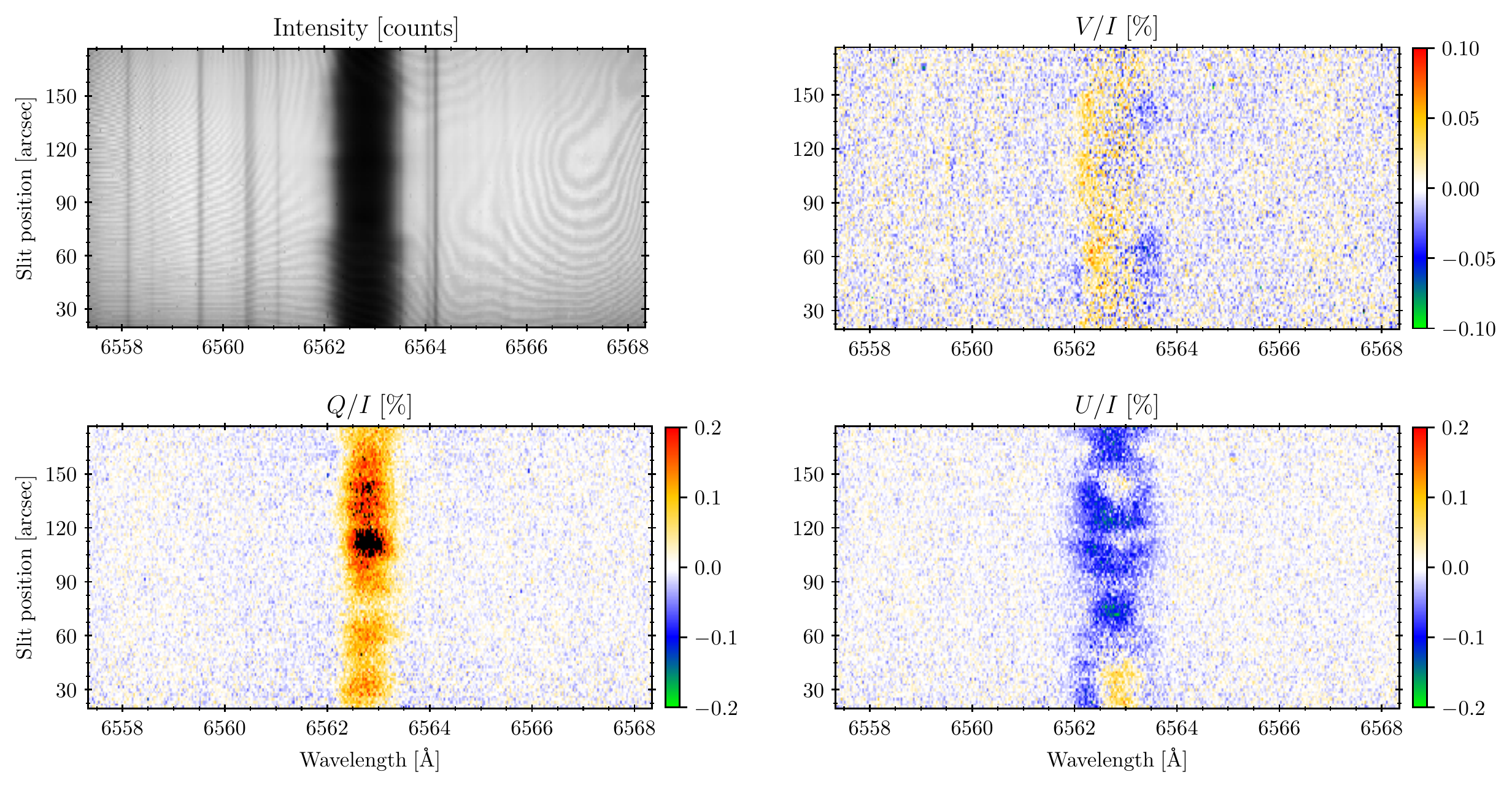}
\caption{Spectropolarimetric $I$, $V/I$, $Q/I$ and $U/I$ images of th
  observation with ID 01. The color scale saturates in black at
  $\pm 0.2~\%$ for $Q/I$ and $UI/$, and $\pm 0.1~\%$ for $V/I$.}
  \label{fig:maps2d_app_id01}
\end{figure*}

\begin{figure*}[h!] \centering
\includegraphics[width=0.96\linewidth]{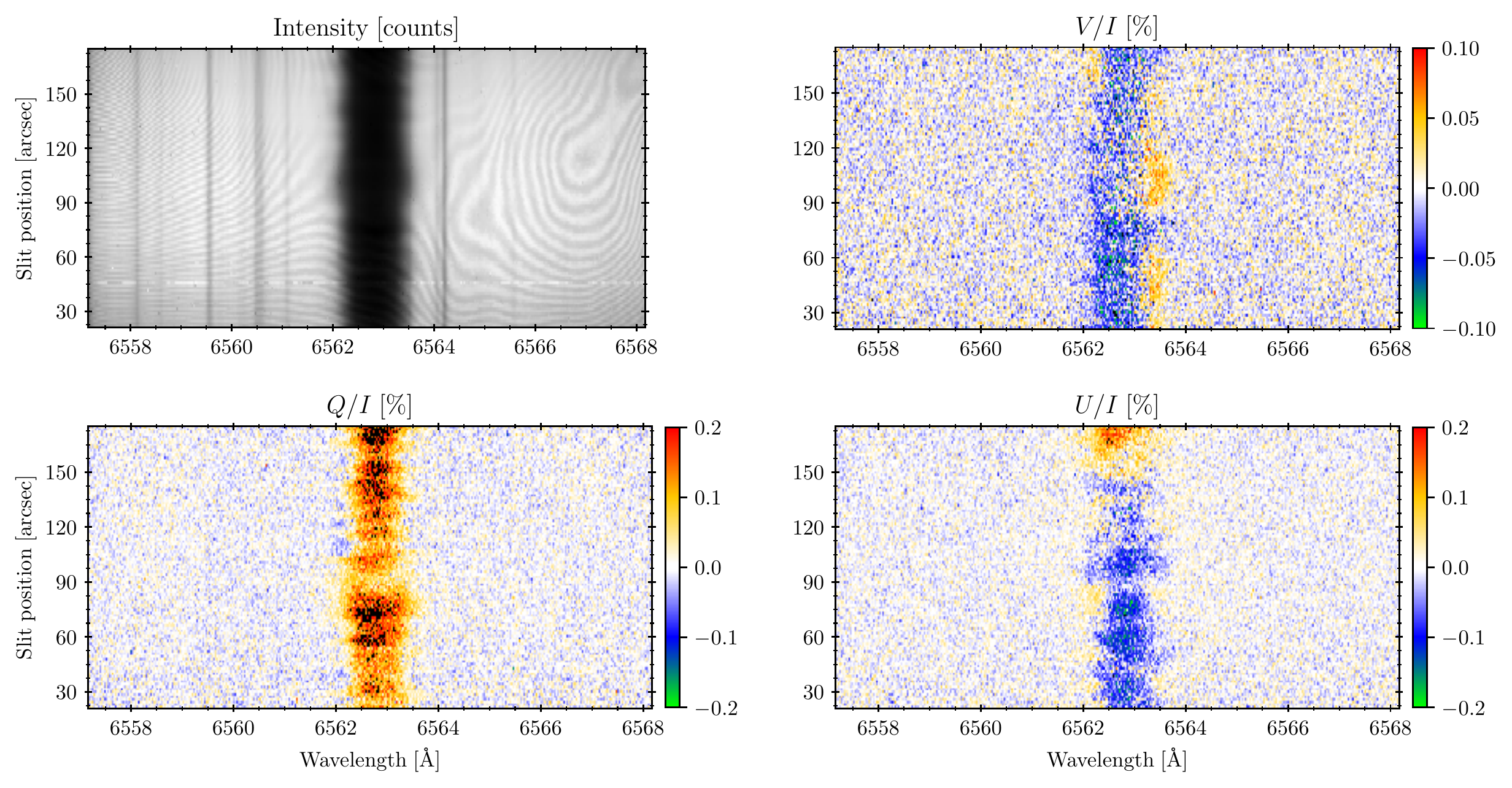}
\caption{Spectropolarimetric $I$, $V/I$, $Q/I$ and $U/I$ images of th
  observation with ID 09. The color scale saturates in black
  at $\pm 0.2~\%$ for $Q/I$ and $UI/$, and $\pm 0.1~\%$ for $V/I$.}
  \label{fig:maps2d_app_id09}
\end{figure*}

\begin{figure*}[h!] \centering
\includegraphics[width=0.96\linewidth]{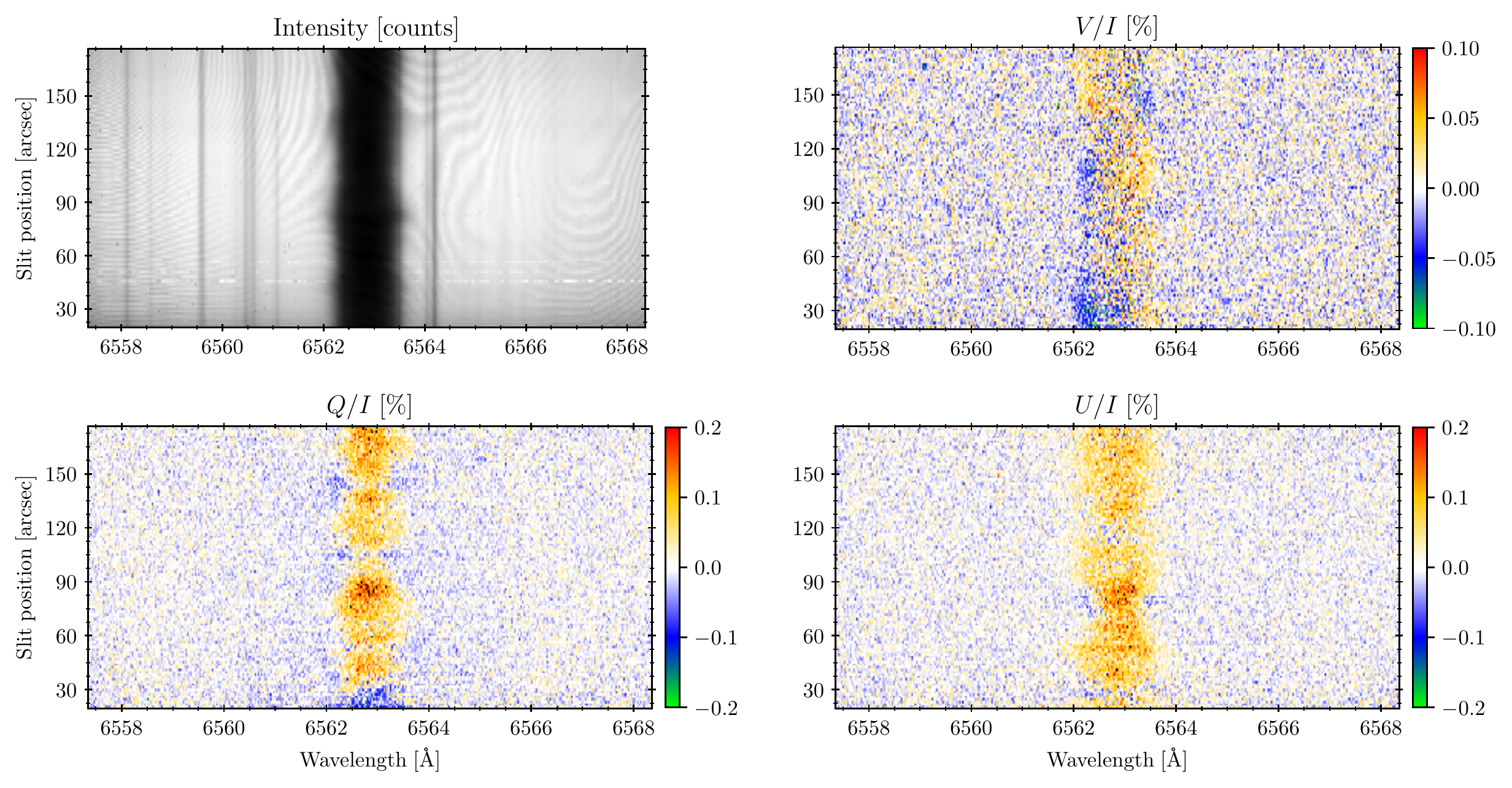}
\caption{Spectropolarimetric $I$, $V/I$, $Q/I$ and $U/I$ images of th
  observation with ID 12. The color scale saturates in black
  at $\pm 0.2~\%$ for $Q/I$ and $UI/$, and $\pm 0.1~\%$ for $V/I$.}
  \label{fig:maps2d_app_id12}
\end{figure*}

\begin{figure*}[h!] \centering
\includegraphics[width=0.96\linewidth]{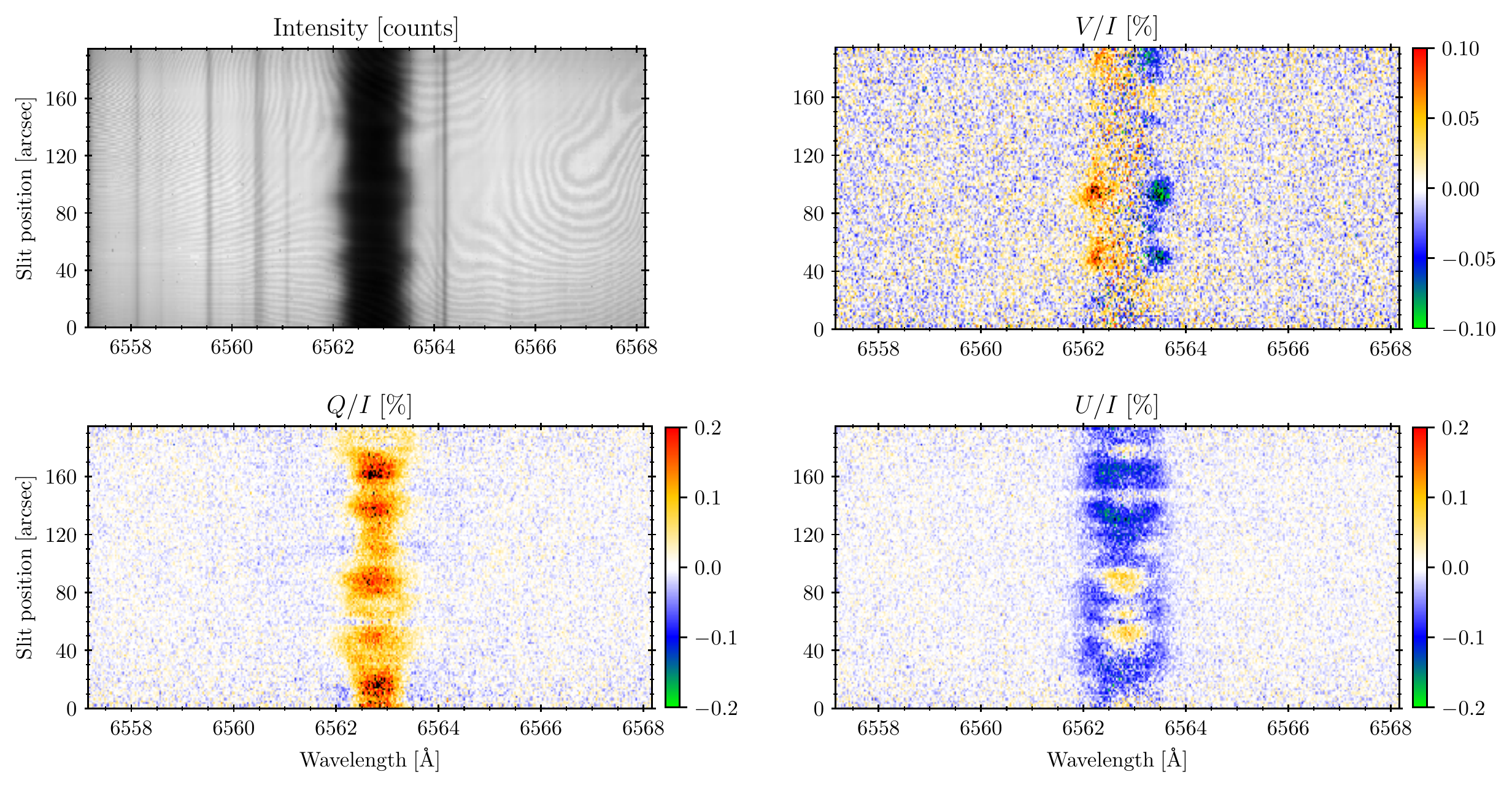}
\caption{Spectropolarimetric $I$, $V/I$, $Q/I$ and $U/I$ images of th
  observation with ID 06. The color scale saturates in black
  at $\pm 0.2~\%$ for $Q/I$ and $U/I$, and $\pm 0.1~\%$ for $V/I$.}
  \label{fig:maps2d_app_north}
\end{figure*}
\begin{figure*}[h!] \centering
\includegraphics[width=0.96\linewidth]{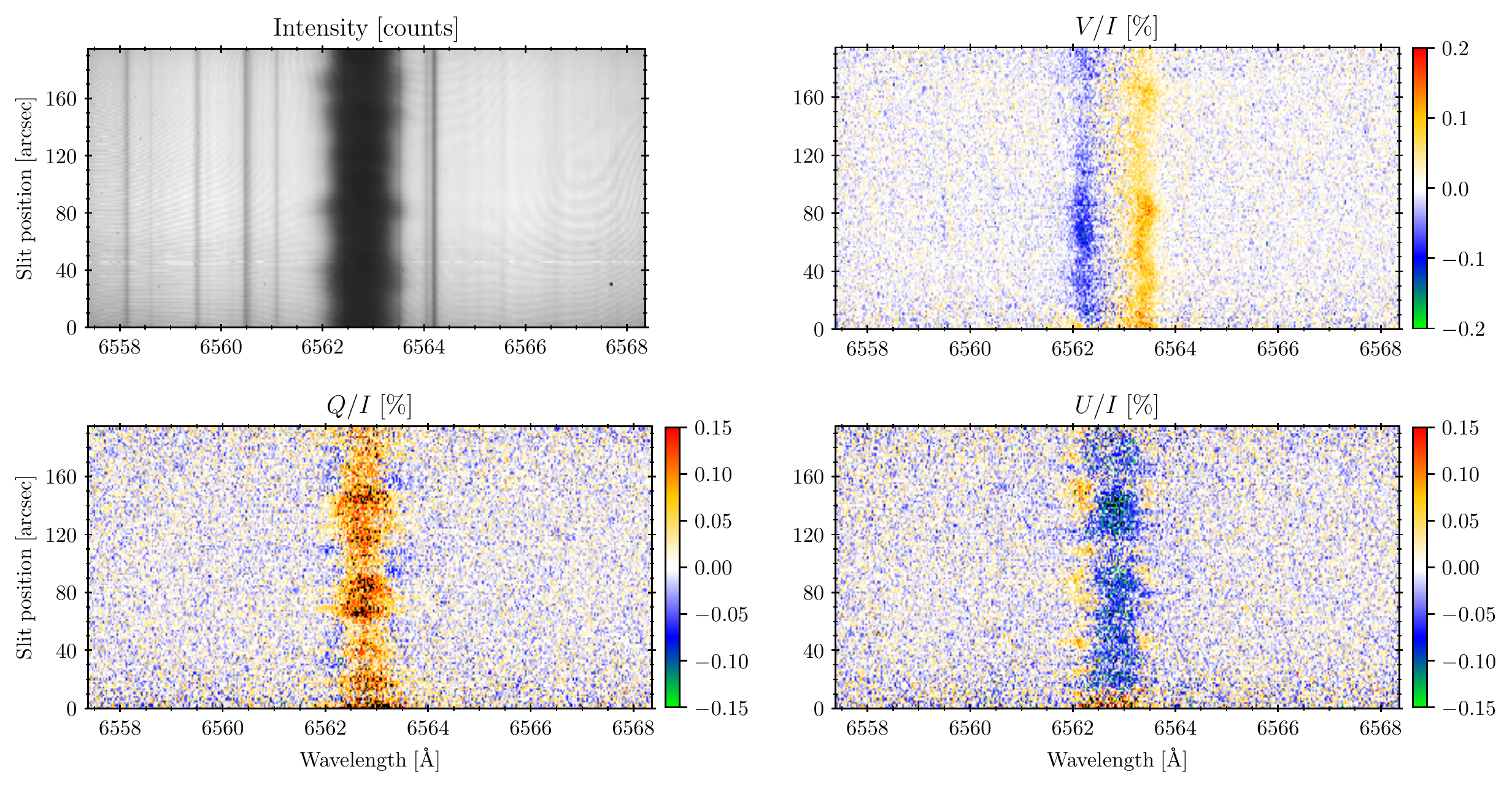}
\caption{Spectropolarimetric $I$, $V/I$, $Q/I$ and $U/I$ images of th
  observation with ID 22. The color scale saturates in black
  at $\pm 0.15~\%$ for $Q/I$ and $UI/$, and $\pm 0.2~\%$ for $V/I$.}
  \label{fig:maps2d_app_east}
\end{figure*}

\begin{figure*}[h!] \centering
\includegraphics[width=0.96\linewidth]{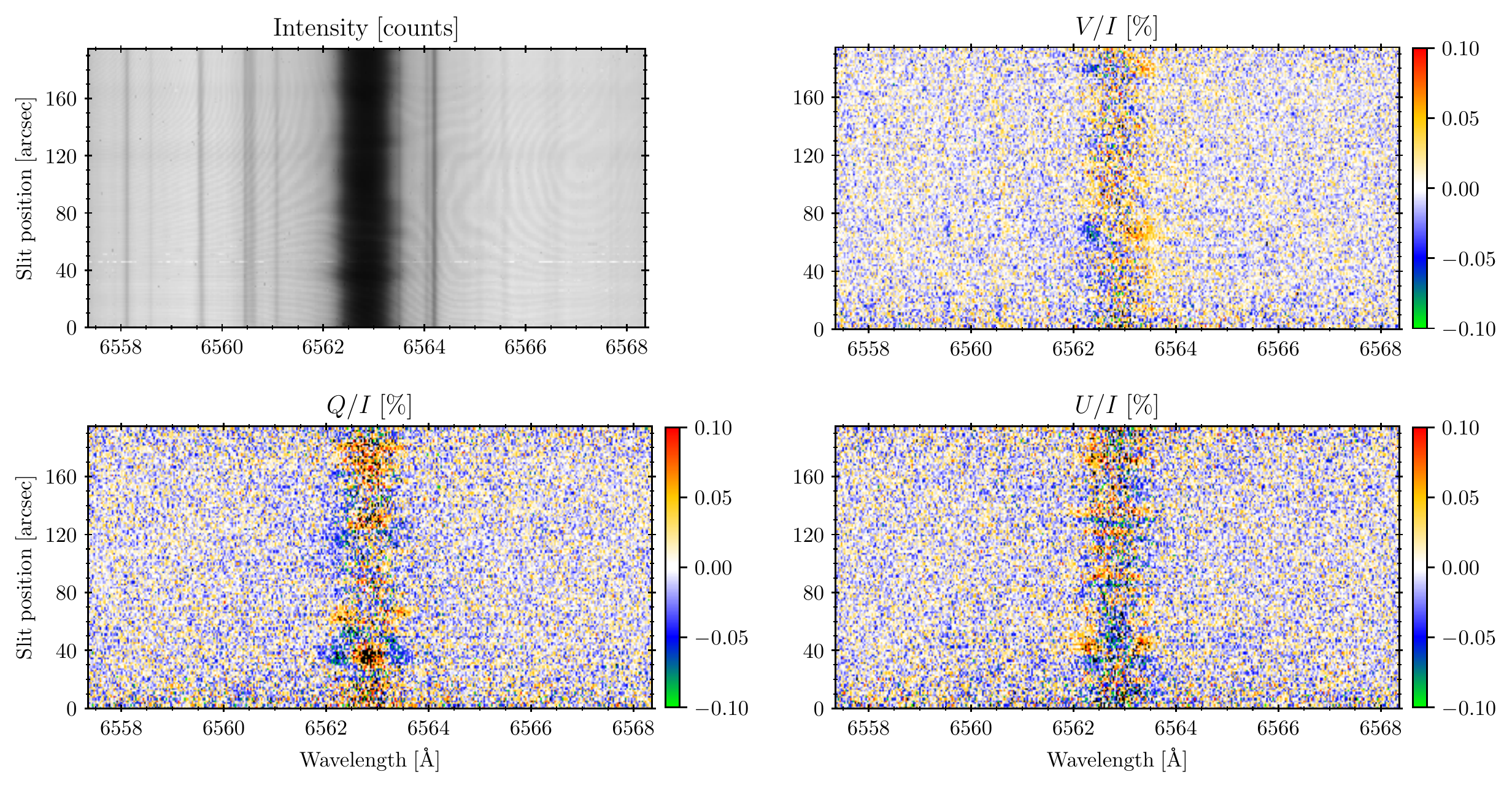}
\caption{Spectropolarimetric $I$, $V/I$, $Q/I$ and $U/I$ images of th
  observation with ID 18. The color scale saturates in black
  at $\pm 0.1~\%$ for $Q/I$, $U/I$ and $V/I$.}
  \label{fig:maps2d_app_west}
\end{figure*}

\begin{figure*}[h!] \centering
\includegraphics[width=0.96\linewidth]{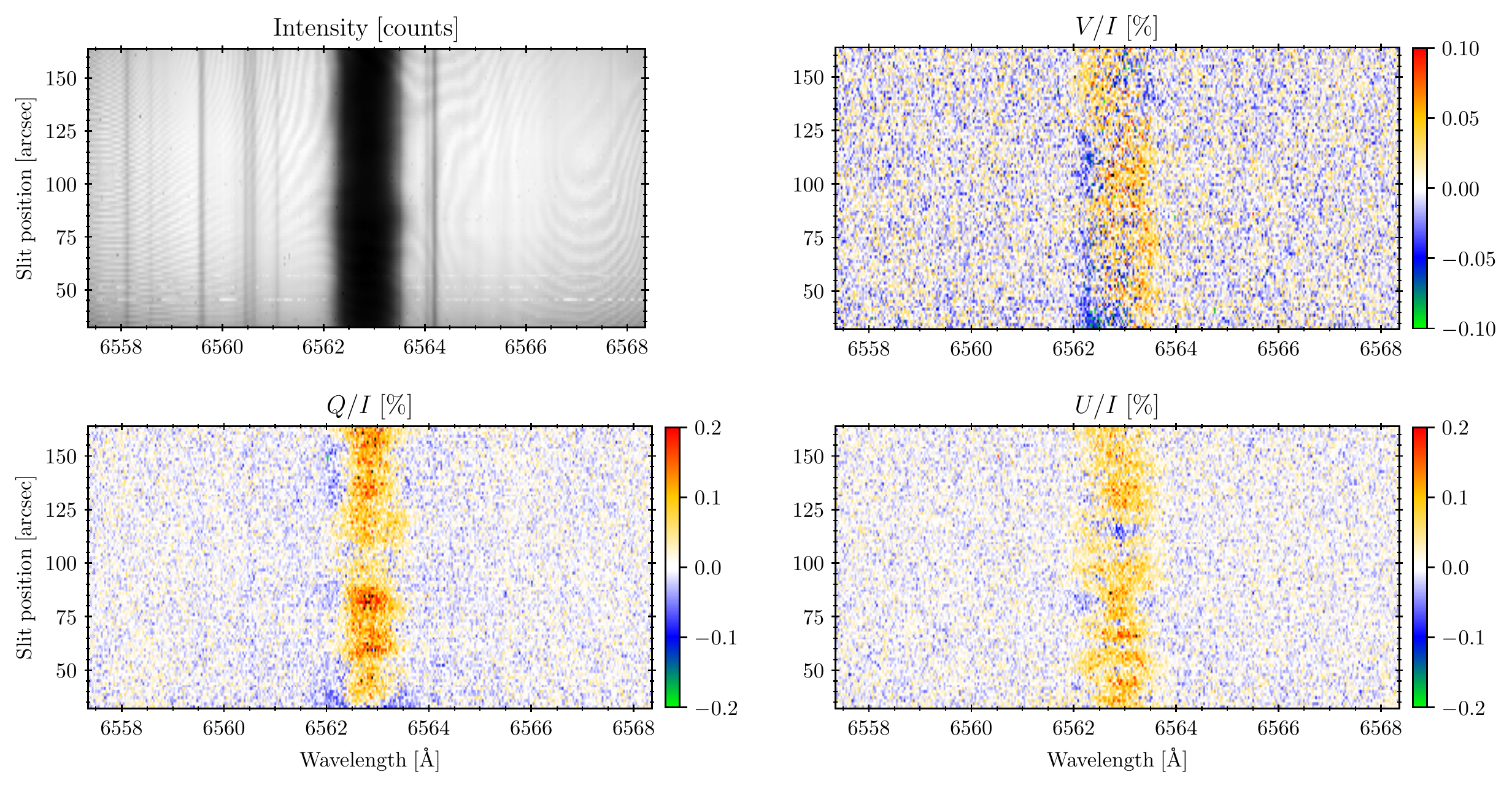}
\caption{Spectropolarimetric $I$, $V/I$, $Q/I$ and $U/I$ images of th
  observation with ID 11. The color scale saturates in black
  at $\pm 0.2~\%$ for $Q/I$ and $UI/$, and $\pm 0.1~\%$ for $V/I$.}
  \label{fig:maps2d_app_id11}
\end{figure*}

\end{appendix}



\end{document}